\documentclass[12pt]{article}
\usepackage[english]{babel}
\usepackage{graphicx,epsfig}
\usepackage{cite}
\usepackage{mathrsfs,amsmath}
\usepackage{slashed}
\usepackage{psfrag}

\newcommand{\be}{\begin{equation}}
\newcommand{\ee}{\end{equation}}
\newcommand{\bea}{\begin{eqnarray}}
\newcommand{\eea}{\end{eqnarray}}

\newcommand{\ba}{\begin{array}}
\newcommand{\ea}{\end{array}}
\newcommand{\balg}{\begin{align}}
\newcommand{\ealg}{\end{align}}

\newcommand{\gsim}{\lower .75ex \hbox{$\sim$} \llap{\raise .27ex \hbox{$>$}} }
\newcommand{\lsim}{\lower .75ex \hbox{$\sim$} \llap{\raise .27ex \hbox{$<$}} }

\begin{document} 

\begin{titlepage}
\vspace*{-1cm}
\flushright{FTUAM-08-26\\ IFT-UAM/CSIC-08-94\\ DFPD-09-TH-01\\ULB-TH/09-01\\IFIC/09-01}
\vskip 2.0cm
\begin{center}
\vspace{0.5cm}
{\Large\bf  Dark coupling }
\end{center}
\vskip 1.0  cm
\begin{center}
{\large M.B. Gavela}$\,^a$~\footnote{\tiny belen.gavela@uam.es,} 
{\large D. Hern\'andez}$\,^a$~\footnote{\tiny d.hernandez@uam.es},
{\large L. Lopez Honorez}$\,^{a,b}$~\footnote{\tiny laura.lopez@uam.es}\\
\vskip .1cm
{\large O. Mena}$\,^c$~\footnote{\tiny mena@ific.uv.es},
and 
{\large S.~Rigolin}$\,^d$~\footnote{\tiny stefano.rigolin@pd.infn.it}\\
\vskip .2cm
\tiny $^a\,$  {Departamento de F\'\i sica Te\'orica and Instituto de F\'\i sica Te\'orica,}
\\
Universidad Aut\'onoma de Madrid, 28049 Cantoblanco, Madrid, Spain
\vskip .1cm
$^b\,$ Service de Physique Th\'eorique,\\
Universit\'e Libre de Bruxelles, 1050 Brussels, Belgium
\vskip .1cm
$^c\,$ Instituto de F\'{\i}sica Corpuscular, IFIC, CSIC and
Universidad de Valencia, Spain
\vskip .1cm
$^d\,$ Dipartimento d Fisica ``Galileo Galilei'', Universit\'a di Padova, via Marzolo 8,
                        I-35131 Padova, Italia
\end{center}
\vskip 0.5cm
\begin{abstract}
The two dark sectors of the universe - dark matter and dark energy - may interact with each other. 
Background and linear density perturbation evolution equations are developed for a generic coupling.
We then establish  the general conditions necessary to obtain models free from early time 
non-adiabatic instabilities. As an application, we consider a viable universe in which the interaction 
strength is proportional to the dark energy density.
The scenario does not exhibit ``phantom crossing" and  is free from instabilities, including 
early ones. A sizeable interaction strength is compatible with combined WMAP, HST, SN, LSS
and $H(z)$ data.
Neutrino mass and/or cosmic curvature are allowed to be larger than in 
non-interacting models. Our analysis sheds light as well on unstable scenarios previously proposed.
\noindent
\end{abstract}
\end{titlepage}

 \newpage
%

\section{Introduction}

Current cosmological measurements point to a \emph{flat} universe whose mass-energy 
composition includes $5\%$ ordinary matter and $22\%$ non-baryonic dark matter, while 
it is dominated by the so-called  ``dark energy'' component, identified as the engine for accelerated  expansion~\cite{Dunkley:2008ie,Komatsu:2008hk,Kowalski:2008ez,Tegmark:2006az,Percival:2006gt}. 

The most economical description of current cosmological measurements explains the nature of dark energy 
as  a Cosmological Constant (CC) in Einstein's equations, representing an invariable vacuum energy 
density. The equation of state $w$ of the dark energy component in the CC case is constant and 
$w = P_{de}/\rho_{de} = -1$, where $P_{de}$ and $\rho_{de}$ denote dark energy pressure 
and density, respectively. However, when computing the vacuum energy density from the quantum 
field theory approach, the naively expected value exceeds the measured one by 123 orders of magnitude 
and it needs to be cancelled by extreme fine-tuning. This unhappy situation has been dubbed the 
CC Problem. Disregarding anthropic justifications, and in the absence of a 
fundamental symmetry which sets the vacuum energy to vanishingly small values, it is appropriate 
to look for alternative physical mechanisms.  A related problem is the so called \emph{why now?} 
or \emph{coincidence} problem, i.e. why the dark matter and dark energy contributions to the 
energy budget of the universe are similar at this precise moment of the cosmic history.

An {\it a priori} appealing avenue is to look for a dynamical explanation of the accelerated expansion.
Under the inspiration of the idea of inflation, it has become quite popular to consider cosmic 
scalar fields, dubbed {\it quintessence}, which would drive the expansion of the universe
~\cite{Caldwell:1998je,Zlatev:1998tr,Wang:1999fa,Wetterich:1994bg,Peebles:1987ek,Ratra:1987rm}.
However, quintessence models are not better than the CC scenario as regards fine-tuning: 
no symmetry explains the tiny value of the potential at its minimum, which is imposed 
{\it by hand}. In spite of this, it seems worth to study the role that scalar fields may play 
in the evolution of the universe and explore the possibility of a dynamic understanding of the problem.

Cosmic scalar fields, if present, may couple to all other fields in nature.
While the strength of interactions between ordinary matter and the dark energy fields are 
severely constrained by observation~\cite{Carroll:1998zi}, significant interactions within the dark 
sectors itself, {\it i.e.} between dark matter and dark energy, are still allowed and could affect 
significantly the universe evolution. Interacting dark matter-dark energy models have been 
proposed since the 90's (see \cite{Amendola:1999qq} for a complete set of references). 
They were first explored in the context of 
coupled quintessence~\cite{Wetterich:1994bg,Amendola:1999dr,Amendola:1999er}. 
Some quintessence models gain extra motivation for being  particular cases of theories of 
modified gravity (i.e. Brans-Dicke theories). Others have been proposed as a natural explanation 
to the coincidence problem~\cite{Comelli:2003cv}. Finally, it was pointed out~\cite{Das:2005yj} that when dark matter and dark energy interact, 
the system may mimic an effective $w<-1$ naturally: current data still allows for such a possibility. 

In this paper we explore the simple idea that the densities of dark matter and dark energy do not 
evolve independently but coupled, although we will not refer to any particular cosmic field. 
The interaction strength between the two dark sectors will be generically dubbed as 
{\it ``dark coupling''}.
To fix the ideas, the possible interactions can be parametrized~\cite{Valiviita:2008iv} by 
 \begin{eqnarray}
  \label{eq:conservDM}
\nabla_\mu T^\mu_{(dm)\nu} &=&Q \,u_{\nu}^{(dm)}/a~, \\
  \label{eq:conservDE}
\nabla_\mu T^\mu_{(de)\nu} &=&-Q \,u_{\nu}^{(dm)}/a~, 
\end{eqnarray}
with $T^\mu_{(dm)\nu}$ and $T^\mu_{(de)\nu}$ the energy momentum tensors for the dark matter 
and dark energy components, respectively.  The dark matter four velocity $u_{\nu}^{(dm)}$ is 
defined in the synchronous gauge, in terms of the fluid proper velocity $v^i_{(dm)}$, as 
$u^{(dm)}_\nu = a (-1, v^i_{dm})$, where $\mu=0..3$ and $i=1..3$.
The coefficient $Q$ encodes the dark coupling, that is, it stands generically for the 
interaction rate between the two 
dark sectors~\footnote{It is to be noticed that an alternative parametrization, in which the right-hand 
side of Eqs.~(\ref{eq:conservDM}) and  (\ref{eq:conservDE}) would be substituted simply by $Q$ and 
$-Q$, would lead for the choices of $Q$ considered in this work to the same results than those found below, 
as the difference only shows up in the equations for the $\theta_{dm}$ evolution, which are not 
operative in the synchronous gauge, comoving with dark matter.}
\footnote{
Notice as well that more covariantly-written  parametrizations are possible and even desirable, such as 
for instance to assume instead a dark coupling term  $\propto T^\mu_{(i)\mu}u_{\nu}^{(j)}$,  
or $\propto T^\mu_{(i)\nu} u_{\nu}^{(j)}$, with $i,j$ denoting some species, and alike 
phenomenological ansatzs. The qualitative findings to be developed in this paper will also hold for 
such constructions, whose details will be explored elsewhere \cite{newUS}. 
In order to compare with previous work in the literature we will stick in this paper to the formulation 
adopted in Eqs.~(\ref{eq:conservDM}) and (\ref{eq:conservDE}) above.}.

In uncoupled models, any $w\ne -1$ value is tantamount to dynamical dark energy and in consequence  
dark energy density perturbations develop, contrary to the case of pure vacuum energy ($w=-1$). 
The dark interaction, when present, will impinge on the density perturbations. 
Recently, using the results of Ref.~\cite{Kodama:1985bj,Bean:2007ny}, a consistent treatment 
of perturbations in models  of dark coupling has been proposed. It was
pointed out in Ref.~\cite{Valiviita:2008iv} 
that early time instabilities may arise for constant equation of state, $w\ne-1$, driven by the coupling 
terms appearing in the non-adiabatic dark energy pressure perturbations. It was also claimed that 
such instabilities are present no matter how weak is the coupling.

 In this paper:
\begin{itemize}
\item We formalize the evolution equations for a general coupling $Q$, up to first order in linear perturbation theory.
\item
We definitely clarify the origin of the non-adiabatic instabilities, identify the instability regions as a  function of the model parameters and propose the general conditions necessary to avoid them.
\item The above results will be then illustrated within a successful class of models, in which $Q$ is proportional to the dark energy density. 
Present data will be shown to allow for a sizeable interaction strength and to imply weaker cosmological limits  on 
neutrino masses or cosmic curvature with respect to non-interacting scenarios. 
\item Finally, a comparative discussion of the existing literature will  shed light on previously proposed models. 
\end{itemize}

\newpage

\section{Background and linear perturbations}
\label{sec:linear}

Consider a flat universe described by the Friedman-Robertson-Walker (FRW) metric. 
We will work in the synchronous gauge \cite{Lifshitz:1945du,Lifshitz:1963ps} comoving 
with dark matter, {\it i.e.} a gauge in which the dark matter peculiar
velocity vanishes. Focusing on the dark energy and dark matter
evolution and assuming pressureless dark matter $w_{dm} = P_{dm}/\rho_{dm}=0$, it follows from 
Eqs.~(\ref{eq:conservDM}) and (\ref{eq:conservDE}) that the evolution of the background 
energy densities is given by:
\begin{eqnarray}
  \label{eq:EOMm}
  \dot\rho_{dm}+ 3\mathcal{H}\rho_{dm} &=& Q\,,\\
\label{eq:EOMe}
 \dot\rho_{de}+ 3 \mathcal{H}\rho_{de}(1+ w)&=&- Q\,, 
\end{eqnarray}
where  the dot indicates derivative with respect to conformal time $d\tau = dt/a$ and 
$\mathcal{H}= {\dot a}/a$.
The effective  background equations of state for the two fluids are thus given by
 \begin{eqnarray}
  \label{eq:weffDM}
  &w_{dm}^{eff}= -\frac{Q}{3\mathcal{H}\rho_{dm} }\,,\\
  \label{eq:weffDE}
  &w_{de}^{eff}= w+\frac {Q}{3\mathcal{H}\rho_{de} }\,,
  \end{eqnarray}
where $w$ would be the dark-energy equation of state  for a vanishing interaction. 
Equations~(\ref{eq:weffDM}) and (\ref{eq:weffDE}) suggest immediately how the interaction 
 between the two fluids can contribute to the effective value of the dark energy 
equation of state. Furthermore, they show that positive (negative) values of $Q$ 
contribute as an effective negative (positive) pressure in the dark matter background equation. 
This leads to less (more) dark matter in the past than in the uncoupled case and, as a consequence, 
the matter radiation equality will happen later (earlier) on.

Also notice that a universe in accelerated expansion today requires $w<-1/3$ even in the presence 
of a dark coupling. Indeed, the deceleration parameter satisfies:
\begin{equation}
  \label{eq:decel}
q=-  \frac{\dot {\mathcal H} }{{\mathcal H}^2}=\frac12 (1+3\,w\,\Omega_{de})
\end{equation}
either with or without dark coupling. In this equation we neglected the curvature contribution. 

The coupling between the two dark sectors will also affect the evolution of the dark matter 
and dark energy density perturbations, $\delta \rho_{dm}$ and $\delta \rho_{de}$, respectively. 
In the synchronous comoving gauge, metric scalar perturbations are described by the two usual 
fields~\cite{Ma:1995ey} $h(x,\tau)$ and $\eta(x,\tau)$.
Defining $\delta \equiv \delta \rho /\rho$ for the fluid density perturbations, 
$\theta \equiv \partial_i v^i$ for the divergence of the fluid proper velocity $v^i$ and 
 using Eq.~(\ref{eq:conservDM}) for presureless dark matter, it results at first order in perturbation theory:
\begin{eqnarray}
\label{eq:deltam}
\dot\delta_{dm}  & = & -\frac12 \dot h+\delta \left[ Q/\rho_{dm} \right]\,.
\end{eqnarray}
For dark energy, using Eq.~(\ref{eq:conservDE}) it follows that:
\begin{eqnarray}
\label{eq:deltae}
\dot\delta_{de}  & = & -(1+w)(\theta_{de}+\frac12 \dot h)
-3 {\mathcal H}\left(\frac{\delta P_{de}}{\delta \rho_{de}} -w\right) \delta_{de} -\delta \left[ Q/\rho_{de} \right] \,,
\\
\label{eq:thetae}
\dot \theta_{de}  & = & -{\mathcal H}(1-3w)\theta_{de} +\frac{k^2}{ 1+w} \frac{\delta P_{de}}{\delta \rho_{de}}\delta_{de}+
\frac{Q}{\rho_{de}}\theta_{de}\,.
\end{eqnarray}
The initial conditions for the dark matter and dark energy density perturbations will be 
taken in what follows as in Ref~\cite{Ma:1995ey}; in particular we assume
$\delta_{de}=0$ initially. For numerical computations and illustrative plots,
we have used the publicly available \texttt{CAMB}
code~\cite{Lewis:1999bs}, which includes the full
evolution for all species (photons, neutrinos, baryons, dark matter
and dark energy), modifying it to take into account the dark coupling. 

\subsection{The doom factor}

Equations~(\ref{eq:deltam}), (\ref{eq:deltae}) and (\ref{eq:thetae}) are written in the dark matter rest frame. 
However, a priori, the dark energy sound speed, $c_{s\,de}^2=\frac{\delta
  P_{de}}{\delta\rho_{de} }$, is only well known in the rest frame of dark
energy \cite{Bean:2003fb}.  It can be shown~\cite{Bean:2003fb} that:
\begin{equation}
\delta P_{de}= \hat c_{s\,de}^2 \delta \rho_{de} - (\hat c_{s\,de}^2- c_{a\,de}^2) \dot \rho_{de} \frac{\theta_{de}}{k^2}\,,
\label{eq:dpcs_in}
  \end{equation}
where $\hat c_{s\,de}^2$  is the propagation speed of pressure fluctuations in the rest frame of dark
energy and $c_{a\,de}^2= \dot P_{de}/\dot \rho_{de}$ is the so called ``adiabatic sound
speed'', which for constant $w$ satisfies $c_{a\,de}^2=w$. 
In the presence of a dark coupling, from Eqs.~(\ref{eq:EOMe}) and (\ref{eq:dpcs_in}) it follows that
 \begin{eqnarray}
   \label{eq:dpcs}
\frac{\delta P_{de}}{\delta \rho_{de}}
&=&\hat c_{s\,de}^2 + 3(\hat c_{s\,de}^2- c_{a\,de}^2) 
 (1+w^{eff}_{de}) \frac{{\mathcal H}\theta_{de}}{k^2\delta_{de}}\nonumber\\
 &=&\hat c_{s\,de}^2 +3(\hat c_{s\,de}^2- c_{a\,de}^2) 
 (1+w)\left(1 + {\bf d}  \right) \frac{{\mathcal H}\theta_{de}}{k^2\delta_{de}}\,,
\end{eqnarray}
where we define 
\begin{equation}
\label{eq:maldito}
 {\bf d} \equiv \frac{Q}{3\mathcal H\rho_{de}(1+w)}\,,
\end{equation}
where $ {\bf d}$ stands for {\it doom}: we dub it so as it is precisely 
 this extra factor, proportional to the dark coupling $Q$, which may
 induce non-adiabatic instabilities in the 
 evolution of dark energy perturbations. 
Its sign will be determinant, as we are going to show.

Rewriting  Eqs.~(\ref{eq:deltam}), (\ref{eq:deltae}) and (\ref{eq:thetae}) in terms of 
$\hat c_{s\,de}^2$ and ${\bf d}$, we have:
\begin{eqnarray}
\label{eq:deltambis}
\dot\delta_{dm}  & = & -\frac12 \dot h+3 {\mathcal H}(1+w)\,\delta \left[ \frac{\rho_{de}}{\rho_{dm}}\,{\bf d} \right]\,,\\ \nonumber \\  
\label{eq:deltaebis}
\dot\delta_{de}  & = & -(1+w)(\theta_{de}+\frac12 \dot h)
 -3 {\mathcal H} (1+w)\,\delta \left[ {\bf d} \right] \nonumber \\
& \,& 
-3 {\mathcal H}\left(\hat c_{s\,de}^2 -w\right)\left[ \delta_{de} +3
{\mathcal H} (1+w)\left( 1 + {\bf d}\right) \frac{\theta_{de}}{k^2} \right]\,
\,, \\ \nonumber \\
\label{eq:thetaebis}
\dot \theta_{de}  & = & -{\mathcal H}\left(1-3\hat c_{s\,de}^2 -3{\bf d}(\hat c_{s\,de}^2+1)
\right)\theta_{de}+\frac{k^2}{ 1+w}\hat c_{s\,de}^2 \delta_{de}\,,
\end{eqnarray}
where $\delta[\,{\bf d}\,]$ includes $\delta Q$ and $\delta_{de}$ contributions. 

Below, in addition to  $w< -1/3$, the speed of sound $\hat c_{s\,de}^2$ will be assumed positive,  
with $\hat c_{s\,de}^2=1$ in numerical computations.

\subsection{Early time (in)stabilities}

In general, the evolution of the dark energy and the dark matter perturbations are directly coupled. Indeed, 
it is known that, even in the uncoupled case, once $w\neq -1 $ the dark matter and dark energy
perturbation evolution depend on each other~\cite{Weller:2003hw} (see also Sec.~\ref{sec:uncoupl}). 

When the dark fluid linear perturbation equations are combined into second order differential equations, 
they take the generic form~\footnote{We surrender here to the extend habit of expressing first 
order temporal differential equations in terms of conformal time,  $\dot\, = \partial /\partial\tau$, 
while  in second order ones $'\,= \partial/\partial \,a$ is used.}:
\begin{eqnarray}
  \label{eq:ddm}
\delta_{dm}'' &=& A_m\, \frac{\delta_{dm}}{a^2} \,+ \,B_m \, \frac{\delta_{dm}'}{a} \, +
    \, \mathcal F (\rho_i, \delta_{i}, \delta_{i}'; { i \ne dm})\,,\\
  \nonumber\\
  \label{eq:dde}
\delta_{de}'' &=& A_e\, \frac{\delta_{de}}{a^2}\, + \, B_e \, \frac{\delta_{de}'}{a} \, + 
    \,{\mathcal G }(\rho_i, \delta_{i},  \delta_{i}'; { i \ne de})\,,
\end{eqnarray}
where $'=\, \partial/\partial a$ and the function $\mathcal F$ ($\mathcal G$) stores the dependence 
in all variables but $\delta_{dm}$ or $  \delta_{dm}'$ ($\delta_{de}$ or $ \delta_{de}'$).  

The evolution of a perturbation will depend on the relative weight of the three terms in the corresponding 
equation {\it and} on their signs: 
\begin{enumerate}
\item For positive $A$, the $A$ and $B$ terms taken by themselves would induce a rapid growth of 
the perturbation, which may be damped or antidamped (reinforced) depending on whether $B$ is 
negative or positive, respectively~\footnote{Obviously, for $|B|>>|A|$ a negative $B$ would prevent 
the onset of growth for any sign of $A$.}. In particular, for $A$ and $B$ both positive, the solution 
may enter in an exponentially growing, unstable, regime. \label{un}
\item For negative $A$, in contrast, the $A$ and $B$ terms taken alone describe a harmonic oscillator, 
with oscillations damped (antidamped) if $B$ is negative (positive). In the $A,B<0$ regime, the third 
term may plays in fact the leading role. \label{dos}
  \end{enumerate}
It  is worth reviewing the uncoupled scenario in detail before proceeding further: 
in it, dark matter perturbations behave as in case \ref{un} above (with $A>0$ and $B<0$), 
while dark energy ones provide an example of behavior as in case \ref{dos}.


\subsection{Uncoupled case}
\label{sec:uncoupl}

Consider first the growth of dark matter and dark energy density perturbations, in the absence 
of dark coupling ($Q=0$), at large scales (${\mathcal H}/k^2\ll 1$) and early times, when 
$\Omega_{de} \delta_{de}$ can be neglected as $\Omega_{de}\ll \Omega_{i} $ for $i=$ 
all other species. In this regime, 
\begin{eqnarray}
 \label{eq:growthm}
   \delta_{dm}'' &=&\frac{3}{2}
   \Omega_{dm}\frac{\delta_{dm}}{a^2}\,-\,\frac32  \frac{\delta_{dm}'}{a}
   +\mathcal F\, ,\\
\delta_{de}'' &=& -\frac92 (c_{s\,de}^2-w)
\,\frac{\delta_{de}}{a^2}\,-\,(\frac52 - 3 w )\frac{\delta_{de}'}{a}+ \mathcal G\,,
\label{eq:growthe}
 \end{eqnarray}
with the functions $\mathcal{F}$ and $\mathcal{G}$ are given by:
\begin{eqnarray}
\label{eq:F}
\mathcal F&=& \frac32  \sum_{i \ne dm}\left(1 + 3 \frac{\delta
  P_i}{\delta \rho_i}\right)\Omega_i\frac{\delta_i}{a^2} \, ,\cr
\label{eq:G}
\mathcal G&=& \frac{(1+w)}{2} \left[ 3 \sum_{i\ne de}\left(1 + 3 \frac{\delta P_i}{\delta \rho_i}\right)\Omega_i\frac{\delta_i}{a^2} - 2\left( 3 \hat{c}_{s\,de}^2-1 \right)\frac{\delta_{dm}'}{a} \right]\,.
\end{eqnarray}

\begin{figure}[t]
\vspace{-0.1cm}
\begin{center}
\begin{tabular}{cc}
\hspace*{-1.5cm} 
\psfrag{c}[c][c]{\tiny{$m$}}
\includegraphics[width=8.5cm]{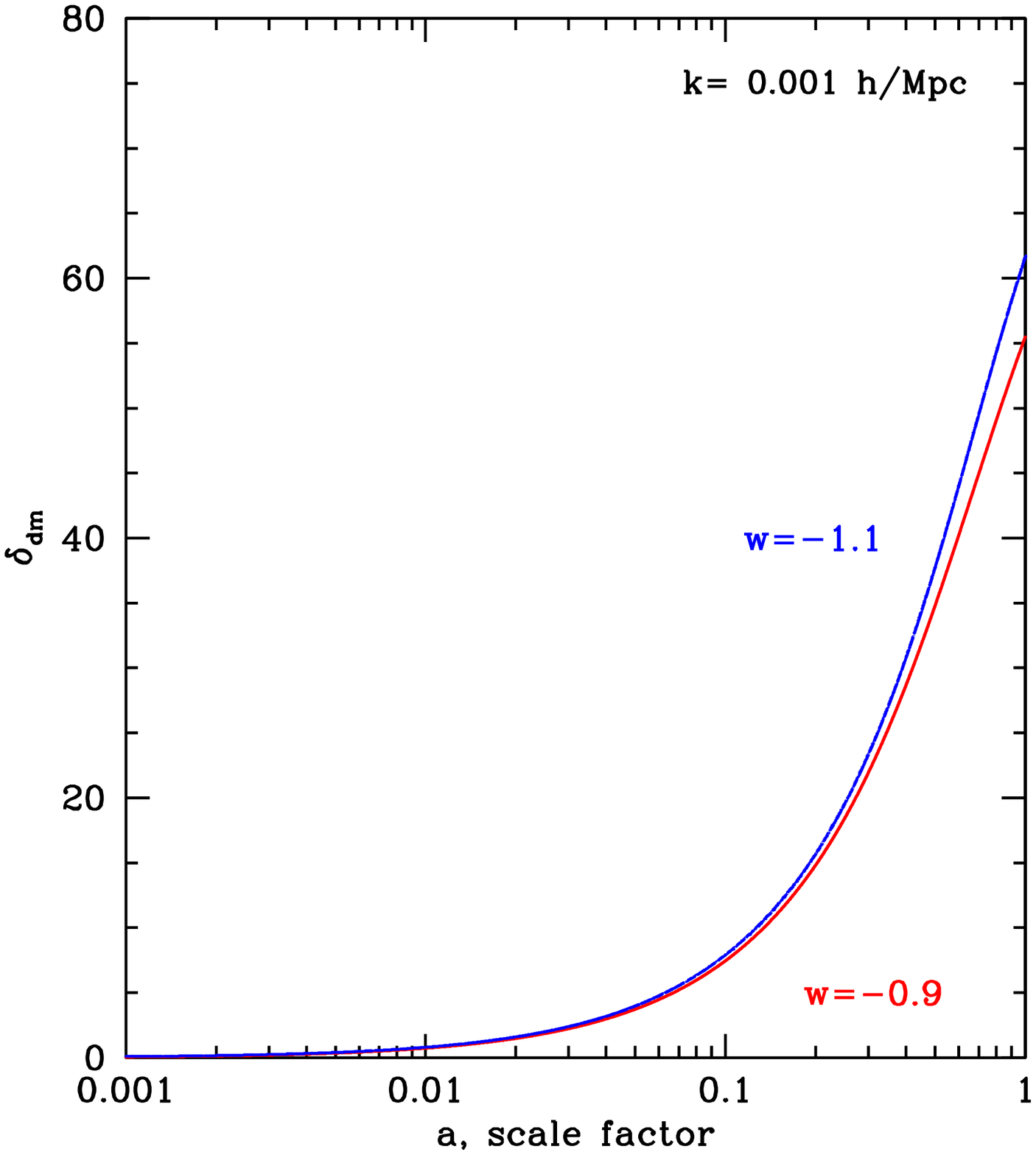} &
		 \includegraphics[width=8.5cm]{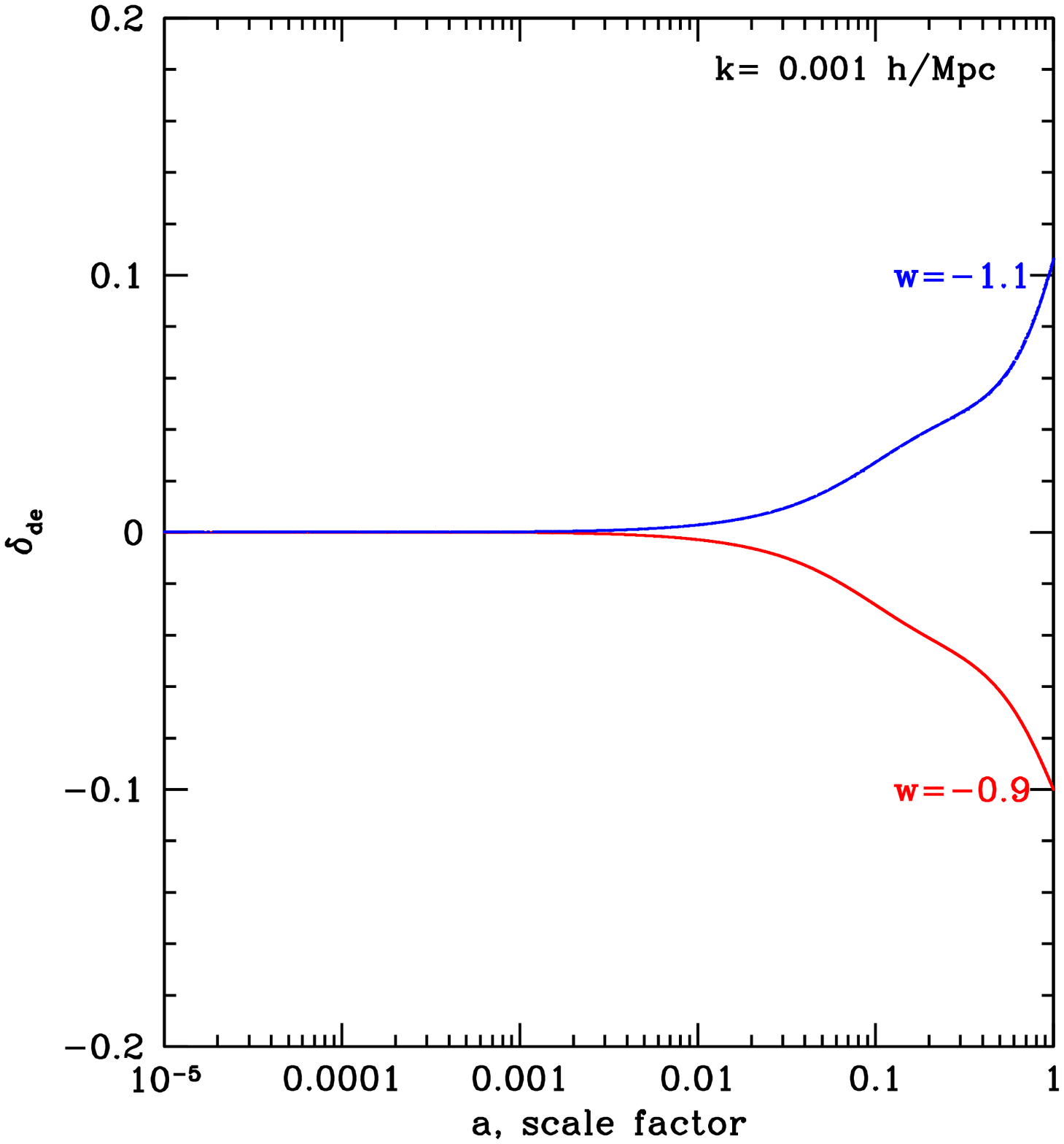} \\
\end{tabular}
\caption{\it  Left panel: The blue (red) curve depicts the
    evolution of the $\delta_{dm}$ perturbation vs the scale factor for 
    $k=0.001$ h/Mpc and $w=-1.1$ ($w=-0.9$). Right panel: same as in
    the left panel but for the $\delta_{de}$ evolution.}
\label{fig:uncoupled}
\end{center}
\end{figure}

Equation~(\ref{eq:growthm}) for dark matter density perturbations has $A>0$ and $B<0$: the latter term
damps the growth of dark matter perturbations propitiated by the $A$ term,
with the overall well known polynomial rising, see Fig.~\ref{fig:uncoupled} (left panel) . 

Equation~(\ref{eq:growthe}) for dark energy perturbations has instead both $A$ and $B$ coefficients
negative, with the first two terms describing then a damped harmonic oscillator and in 
this case the contribution of radiation, dark matter and matter encoded in $\mathcal G$ 
drives the evolution, see Fig.~\ref{fig:uncoupled} (right panel). Equations~(\ref{eq:growthe}) and (\ref{eq:G}) 
also illustrate that, for constant $w\ne-1$, dark energy perturbations do develop even in 
absence of coupling, seeded by the $\mathcal G$ term, in contrast to the $w=-1$ (pure 
vacuum energy) case, in which no perturbation can develop in the dark energy background. 

\subsection{Strongly coupled case}
\label{strongcoupling}

Consider now the opposite case in which the dark-coupling terms dominate over the usual one. 
For $\hat c_{s\,de}^2>0$ considered all through, the {\it strong coupling regime} can be characterized by 
 \begin{eqnarray}
 |{\bf d}|&=&
\left|\frac{Q}{3\mathcal H\rho_{de}(1+w)}\right|\,>\,1 \,,
  \label{eq:condstr}
\end{eqnarray}
which guarantees that the interaction among the two dark sectors drives the non-adiabatic 
contribution to the dark energy pressure wave, see Eq.~(\ref{eq:dpcs}), and becomes the leading term 
in Eq.~(\ref{eq:deltaebis}) for $\dot \delta_{de}$ and Eq.~(\ref{eq:thetaebis}) for $\dot \theta_{de}$. 
 At large scales, those equations  reduce to:
\begin{eqnarray}
\label{eq:deltaxi}
\dot\delta_{de}  & \simeq&
-3 {\mathcal H}\left(\hat c_{s\,de}^2 -w\right)  \left( \delta_{de} +3
{\mathcal H} (1+w) \,{\bf d}\,
\frac{\theta_{de}}{k^2} \right)  -(1+w)\left(
3
{\mathcal H} \,\delta\left[ {\bf d}\right]+\frac{\dot h}{2}\right)\,,
\\
\label{eq:thetaxi}
\dot \theta_{de}  & \simeq& 3
{\mathcal H}\, {\bf d}  \,
  (\hat c_{s\,de}^2+1)\,\theta_{de}+ \frac{k^2\hat c_{s\,de}^2}{1+w}\, \,\delta_{de} \,.
\end{eqnarray}
The complete second order differential equation describing the growth of dark energy perturbations 
can be found in Appendix \ref{a:grstrongfull}; for values of $w$ near $-1$ and in the strong coupling 
regime, the $\delta_{de}$ and $\delta_{de}'$ contribution to the
second order differential equation reads
  \begin{eqnarray}
  \label{eq:grstrongfullours}
\delta_{de}''&\simeq&  \,3\,{\bf d}\,(\hat c_{s\,de}^2+1)\left(\,\frac{\delta_{de}'}{a} 
    \,+\,3\frac{\delta_{de}}{a^2}\frac{(\hat c_{s\,de}^2-w)}{\hat c_{s\,de}^2+1}
     \,+\,\frac {3(1+w)}{a^2}\delta[\,{\bf d}\,]\,\right)+...
  \end{eqnarray}
The sign of the  coefficient $B_e$ of $\delta_{de}^\prime$  in this expression is crucial for 
the analysis of instabilities, as previously argued, see Eq.~(\ref{eq:dde}). Notice that the size 
and sign of the $ B_e$ coefficient also determines the growth rate of $\theta_e$ in 
Eq.~(\ref{eq:thetaxi}).  Assuming $\hat c_{s de}^2>0$, it reduces to the sign of the doom 
factor ${\bf d}$ defined in Eq.~(\ref{eq:maldito}). 
 
Indeed, as previously argued, a positive ${\bf d}$  acts as an antidamping source in the growth 
Eq.~(\ref{eq:grstrongfullours}). Whenever ${\bf d}>1$, it will trigger an exponential runaway 
growth of the dark energy perturbations when simultaneously 
 the overall sign of the $A_e$ coefficient of $\delta_{de}$, resulting from the 
last two terms in Eq.~(\ref{eq:grstrongfullours}), is  also positive.
  Large scale instabilities arise then and the universe appears to be nonviable.


\section{A simple viable model: $Q\propto \rho_{de}$}
\label{ourmodel}

Let us consider now a specific simple coupled model, with the interaction rate $Q=\xi {\mathcal H} 
\rho_{de}$. Equations~(\ref{eq:conservDM}) and (\ref{eq:conservDE}) become consequently:
\begin{eqnarray}
  \label{eq:conservDMxi}
 \nabla_\mu T^\mu_{(dm)\nu} &=&\xi{\mathcal H}\rho_{de} u_{\nu}^{(dm)}/a~, \\
\label{eq:conservDExi}
 \nabla_\mu T^\mu_{(de)\nu} &=&-\xi{\mathcal H}\rho_{de} u_{\nu}^{(dm)}/a~, 
\end{eqnarray}
where $\xi$ is a dimensionless coupling which will be taken as constant. It parametrizes in this 
model the dark coupling strength.
Somewhat similar coupled models have been explored in the literature~\cite{Amendola:1999er,
Amendola:2000uh,Olivares:2006jr,Valiviita:2008iv,CalderaCabral:2008bx,delCampo:2008jx} 
In particular the authors of Ref.~\cite{Valiviita:2008iv}, assuming $Q \propto \rho_{dm}$ or 
$Q\propto \rho_{dm}+\rho_{de}$, concluded that their models, with constant $w$, do 
not provide viable scenarios due to the presence of early time instabilities.

We will show here, though, that the model described by Eqs.~(\ref{eq:conservDMxi}) and (\ref{eq:conservDExi}) 
can satisfy all current observational constraints, without suffering from the instabilities pointed out
in~\cite{Valiviita:2008iv}, even for constant $w$. In this section we will analyze this model in detail,  
postponing, and extending, the discussion of the scenarios presented
in Ref.~\cite{Valiviita:2008iv} 
to Sec.~\ref{sec:Val}. 

\subsection{Background}
The dark matter and dark energy background densities evolve in this model as two fluids,  coupled with a 
strength linearly dependent on the dark energy density present at any given time of the cosmic evolution,
\begin{eqnarray}
  \label{eq:EOMmxi}
  \dot\rho_{dm}+ 3\mathcal{H}\rho_{dm} &=& \xi \mathcal{H} \rho_{de}\\
\label{eq:EOMexi}
 \dot\rho_{de}+ 3 \mathcal{H}\rho_{de}(1+ w_{de})&=&- \xi \mathcal{H} \rho_{de}\,,
\end{eqnarray}
which then correspond to two fluids with effective equations of state given by
 \begin{eqnarray}
  &w_{dm}^{eff}=-\frac{\xi}{3}\frac{\rho_{de}}{\rho_{dm} }\,,\label{weff_dm_I}\\
  &w_{de}^{eff}= w+\frac {\xi}{3 }\,.
  \label{weff_de_II}
  \end{eqnarray}
Note that for constant $w$,  $w_{de}^{eff}$ is also constant while $w_{dm}^{eff}$ is redshift dependent. 
The solutions to equations (\ref{eq:EOMm}) and (\ref{eq:EOMe}) are then: 
\begin{eqnarray}
 \rho_{dm}&=& \rho_{dm}^{(0)} a^{-3} + 
      \rho_{de}^{(0)}\frac{\xi}{3w^{eff}_{de}}(1- a^{-3w^{eff}_{de}})  a^{-3} \label{eq:adm}\,,\\
  \rho_{de}&=& \rho_{de}^{(0)}a^{-3(1+w^{eff}_{de})}\label{eq:ade}\,.
\end{eqnarray}
\begin{figure}[h!]
\begin{center}
\includegraphics[height=.4\textheight]{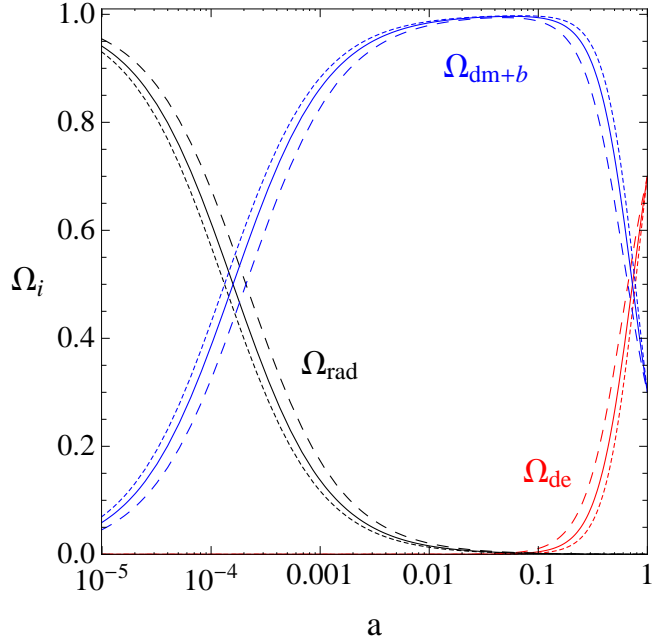}
\caption{\emph{Scenario with $Q \propto \rho_{de}$.
Relative energy densities of dark matter plus baryons $\Omega_{dm+b}$ (blue), radiation $\Omega_{rad}$ (black) 
and dark energy $\Omega_{de}$ (red), as a function of the scale factor $a$, for w=-0.9. 
Three values of the coupling are illustrated: $\xi= 0$ (solid curve), $0.25$ (long dashed curve) and
$-0.25$ (short dashed curve). }}
  \label{fig:fishesplots}
\end{center}
\end{figure}
The dark energy density is thus always positive, all along the cosmic evolution and since its 
initial moment. To ensure that the same happens with the dark matter density, all values of 
$w<0$ are acceptable for $\xi<0$, while for positive $\xi$ it is required that $\xi\lsim -w$. 

The resulting evolution of the relative energy densities is shown in Fig. \ref{fig:fishesplots}, where
  the present matter and dark energy densities have been imposed to be $0.27$ and $0.73$ (a procedure to be repeated all through the paper).
   We see that negative (positive) couplings lead to more (less) dark matter in the past than in
the uncoupled case, as expected. As a consequence, the matter radiation equality happens
earlier (later) on. Fig.~(\ref{fig:fishesplots}) illustrates as well that positive (negative) values of 
$Q$ soften (worsen) the coincidence problem, although the difference turns out 
to be quantitatively minor for the phenomenologically allowed values of $\xi$. It will be shown 
in Sect.~\ref{sec:Olga} that models with $\xi<0$ and $1+w>0$ will give the best agreement 
with large scale structures observations.

Finally, Eqs.~(\ref{weff_de_II}) and (\ref{eq:ade}) show that the so-called phantom regime, in which 
the dark energy density would diverge in the future, can only happen in this model for 
$1+w^{eff}_{de}<0$, that is, for $w < -1-\xi/3$, instead of $w<-1$ in the uncoupled case. 

\subsection{Dark coupling vs. dynamical dark energy}

From the previous analysis 
 one can wonder what would be the dark energy equation of state reconstructed from 
observational data, if analyzed assuming no coupling. Indeed a dynamical, redshift-dependent, 
equation of state $\tilde w(z)$ (where $z$ denotes redshift), can be mimicked by the combination 
of constant $w$ plus the dark coupling $\xi$ .  

A wealth of data are sensitive to the Hubble parameter $H(z)$ or functions of its integral.  
When such data are analyzed assuming no dark coupling, the following expression for the 
Friedmann equation is to be considered~\footnote{We obviate here as well, for the sake 
of the argument, the contribution to $\Omega^{(0)}_{TOT}$ of any species other 
than dark matter and dark energy: baryons, neutrinos and curvature.}:
\begin{equation}
  \label{eq:RHwz}
R_H(z)=\frac{H^2(z)}{H_0^2}= \Omega^{(0)}_{dm} (1+z)^3 +\Omega^{(0)}_{de} 
\exp\left[3\int_0^z dz'\frac{1+ \tilde w(z')}{1+z'}\right]\,,
\end{equation}
where $\Omega^{(0)}_{i} = \rho^{(0)}_i/\rho^{(0)}_c$, being $\rho^{(0)}_c$ the critical energy 
density today. Relation (\ref{eq:RHwz}) can be inverted to obtain $\tilde w(z)$ as a function of $R_H(z)$:
\begin{equation}
  \label{eq:wH}
 \tilde w(z)=\frac13 \;\frac{ (1+z) dR_H/dz -3 R_H  }{R_H -\Omega^{(0)}_{dm} (1+z)^3 }\, .
\end{equation}
An analogous relation between the redshift dependent equation of state and the luminosity distance 
and its derivative was also obtained previously~\cite{Clarkson:2007bc}.

In contrast, in the presence of the dark coupling, from Eqs.~(\ref{eq:adm}) and~(\ref{eq:ade}) 
it results for the simple model analyzed in this section 
\begin{eqnarray}
R_H(z)&=& (1+z)^{3}\left[ \Omega_{dm}^{(0)}+ \Omega_{de}^{(0)}\frac{\xi}{3w^{eff}_{de}}
\left( 1-  (1+z)^{3 w^{eff}_{de}}\right) + \Omega_{de}^{(0)}(1+z)^{3w^{eff}_{de}}\right]\,.
\label{eq:RH}
\end{eqnarray}
Comparing Eqs.~(\ref{eq:RHwz}) and (\ref{eq:RH}),  it follows the relation between the hypothetically 
reconstructed expression of a dynamical equation of state $\tilde w(z)$ and  the constant $w,\xi$ 
parameters of the darkly coupled universe:
 \begin{equation}
   \label{eq:wzdelta}
\tilde w(z)=\frac{w}{1-\frac{\xi}{3w^{eff}_{de}}(1- (1+z)^{-3w^{eff}_{de}})}\,,
 \end{equation}
 an expression which at small redshifts tends to: 
 \begin{equation}
 \tilde w(z)\sim w(1 + \xi\,z)\,,
 \end{equation}
while $\tilde w(z)\rightarrow 0$ at very large redshifts. 

A striking implication of Eq.~(\ref{eq:wzdelta}) is that the reconstructed equation of state 
would show a peculiar divergent behavior for those models in which the denominator vanishes.
In fact, for modified gravity models and coupled scalar-tensor models
 a similar behavior was 
previously pointed out \cite{Amendola:2007nt,Tsujikawa:2008uc}: their reconstructed $\tilde w(z)$ 
could appear singular at some point in the past and cross the ``phantom boundary'' $w=-1$. 

In the phenomenologically viable model discussed here {\it no} such divergent behavior can arise in the $\tilde w(z)$ 
reconstruction, for negative values of the dark coupling $\xi$ for any $w<0$. Whereas, for 
positive dark coupling values,  a ``phantom-like'' behavior appears, with the divergence 
occurring at redshift $z\sim {\mathcal O} (1-10)$, depending on the specific values of 
the parameters. 

\begin{figure}[t!]
  \begin{center}
\includegraphics[height=.35\textheight]{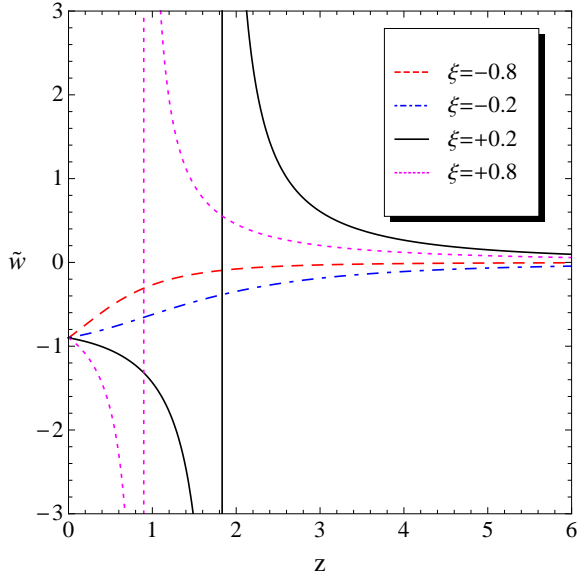}
\caption{\emph{ Scenario with $ Q\propto \mathcal \rho_{de}$. Reconstructed $\tilde w(z)$ 
as function of $z$, for $w= -0.9$. The black (solid) and magenta (short dashed) 
curves depict the $\tilde w(z)$ behaviour for $\xi=0.2$ and $\xi= 0.8$. The blue (long-short 
dashed) and the red (long dashed) curves denote the $\tilde w(z)$ behaviour for $\xi=-0.2$ 
and $\xi=-0.8$. Notice that for positive values of $\xi$ we recover the divergent phantom-crossing 
behavior appearing in scalar-tensor theories.}}
  \label{tab:wzrecons}
\end{center} 
\end{figure}

All these different behaviors are illustrated in Fig.~\ref{tab:wzrecons}, for $w=-0.9$ and 
$\xi=\pm 0.8,\pm 0.2$. The model with $w= -0.9$ and $ \xi = 0.8$ is not viable, due to 
negative dark matter energy density, and it is shown only for illustration purposes.
Notice that an apparently diverging behavior of the equation of state could thus be a clean indicator 
of  the sign of a putative dark coupling. Furthermore, the degeneracy between 
a dynamical equation of state and a constant $w$ plus dark coupling, discussed in this subsection, 
can be disentangled with other $z-$sensitive data, such as precision BAO or SuperNovae measurements 
and the growth of perturbations seeding large structure formation, as it will be seen in Sec.~\ref{sec:Olga}.

\subsection{Linear perturbation theory}

The propagation of dark energy pressure waves is driven by Eq.~(\ref{eq:dpcs}) as explained above, 
with the doom factor Eq.~(\ref{eq:maldito}) given in this case by:
 \begin{equation}
\label{eq:maldito_us}
{\bf d }= \frac{\xi}{3(1+w)}\,.
\end{equation}
The evolution of the perturbations, in the synchronous comoving gauge, is described by 
Eqs.~(\ref{eq:deltambis}), (\ref{eq:deltaebis}) and (\ref{eq:thetaebis}) with, in the present scenario,
\begin{eqnarray}
&\,& \delta\left[{\bf d}\right]=0\,,\label{eq:deltad_us} \\
&\,& \delta\left[\frac{\rho_{de}}{\rho_{dm}}\,{\bf d}\right]= \,{\bf d}\, \frac{\rho_{de}}{\rho_{dm}}(\delta_{de}-\delta_{dm})\,.\label{eq:deltak_us}
\end{eqnarray}

The  dark matter density perturbation changes in size with respect to the uncoupled case.
It is easy to confirm numerically that the major effect of the coupling on the dark matter distribution 
results from the background, though, for dark coupling values small enough so as to allow viable models.
For negative couplings, there is more dark matter in the past. As a consequence, dark matter 
perturbations cluster more and the density perturbation is larger.

\subsection{Early time (in)stability}
\label{sec:instab}

\begin{figure}[htb]
\vspace{-0.1cm}
\begin{center}
\begin{tabular}{cc}
\hspace*{-0.75cm} 
\includegraphics[width=8cm]{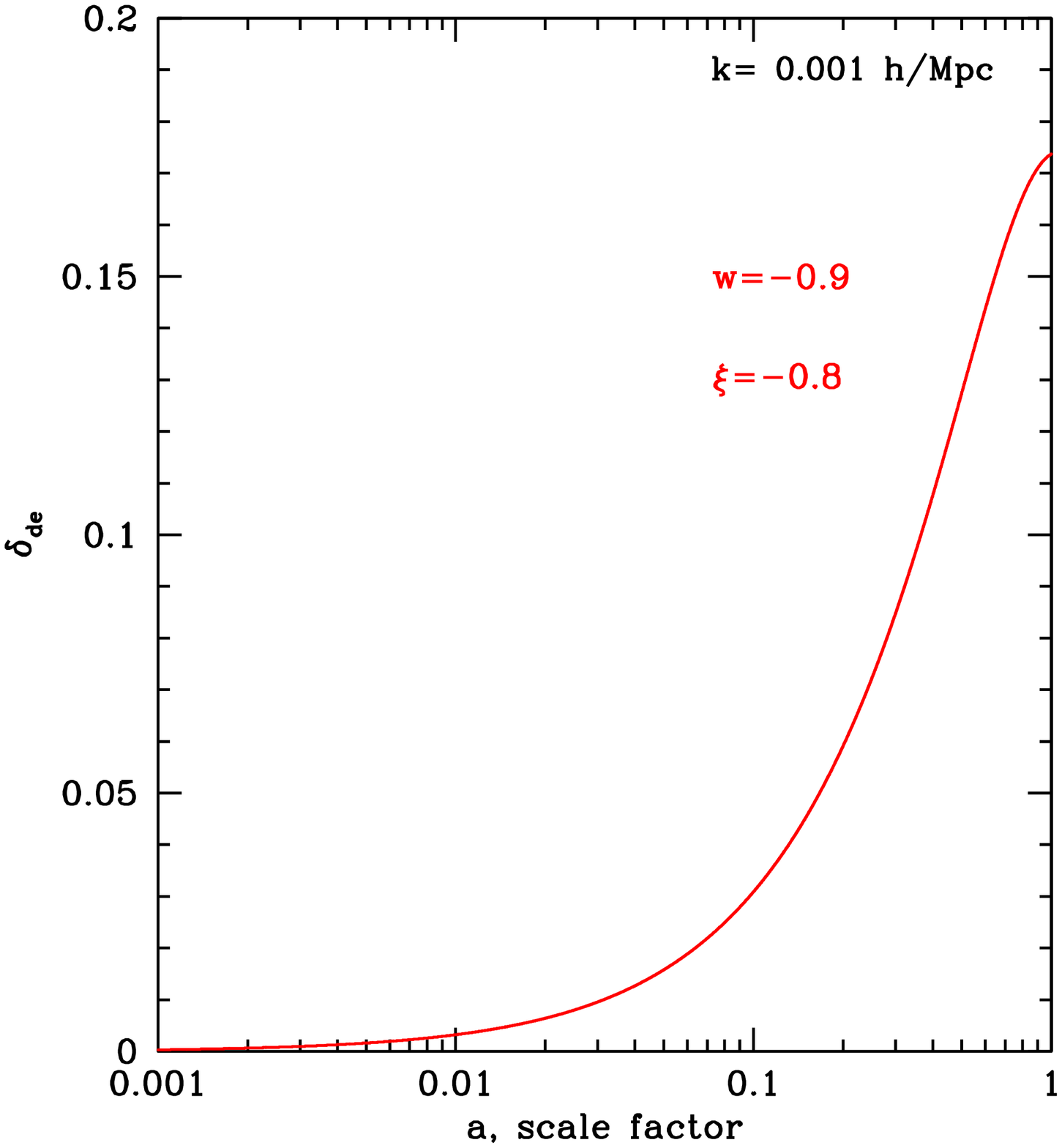} &
\includegraphics[width=8cm]{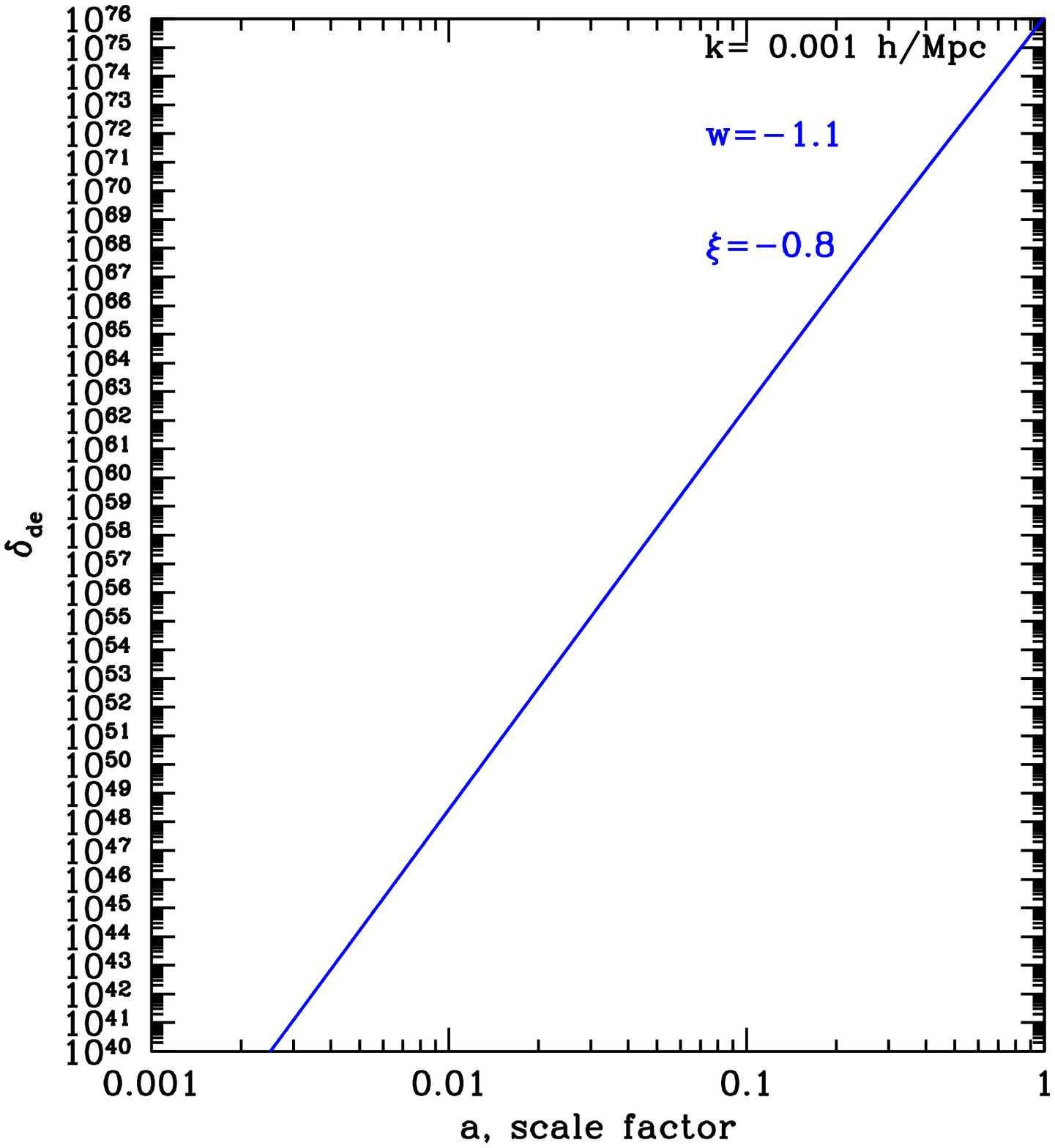} \\
\end{tabular}
\caption{\it Scenario with $Q\propto \rho_{de}$ in the strong coupling
  regime. Left (right) panel: Evolution of the $\delta_{de}$ perturbation
 vs the scale factor for scales $k=0.001$ h/Mpc, for $\xi=-0.8$ and 
 $w=-0.9$ ($w=-1.1$). Notice the  early time instability present when the
 doom factor ${\bf d }$ in Eq.~(\ref{eq:maldito_us})
  is sizeable and positive, as predicted from the 
 study of the instabilities.}
\label{fig:figinstability}
\end{center}
\end{figure}

The strong coupling regime is defined by $|{\bf d}|>1$, as explained in Sec.~\ref{strongcoupling}.
 In this regime and for large scales, $\mathcal H/k \gg 1$, the dark energy perturbations are described by Eqs.~(\ref{eq:deltaxi}) 
and (\ref{eq:thetaxi}), where in this case Eqs.~(\ref{eq:maldito_us}), (\ref{eq:deltad_us}) 
and (\ref{eq:deltak_us}) above apply. It leads, for constant $w\ne-1$,
to a $\delta_{de}$ and $\delta_{de}'$ contribution to the early time growth at large scales, described by

\begin{eqnarray}
  \label{eq:grstrongfull}
  \delta_{de}''&\simeq& 3(1+\hat c_{s\,de}^2)\,{\bf d }\,
  \left(\,\frac{\delta_{de}'}{a} \,+\,3\frac{\delta_{de}}{a^2}
  \frac{(\hat c_{s\,de}^2-w)}{\hat c_{s\,de}^2+1}\,\,\right)+... 
  \end{eqnarray}
which shows that the sign of ${\bf d }$ defines the (un)stable regimes~\footnote{Recall that 
$\hat{c}_{s\,de}^2\ge 0$ is assumed all through.}: 
\begin{enumerate}
\item For  ${\bf d } <0$, that is, for $\xi<0$ and $1+w>0$ (or $\xi>0$ and $1+w<0$), no instabilities are expected;
\item When $\xi$ and $1+w$ have the same sign, instabilities can
  develop at early times whenever ${\bf d } >1$ .
\end{enumerate}
Sweeping over all possible values of $\xi$ and $w$, we have indeed confirmed numerically these predictions.
We summarize our instability criteria for the present model  in Tab.~\ref{tab:rhoe}. An example of the 
onset of unstable behaviour is shown in Fig.~\ref{fig:figinstability}, right panel, in which the particular
choice of $w$ and $\xi$ makes the model unstable at early times, see Tab.~1. The left panel of Fig.~\ref{fig:figinstability} depicts instead the dark energy perturbation, $\delta_{de}$, versus the 
scale factor $a$ in a model free of instabilities and pathological
behaviors in the linear perturbation evolution, notwithstanding the strong coupling regime.
\begin{table}[bt!]
\begin{center}
\begin{tabular}{|c||c|c|c|c|c|c|} \hline
Model: $ Q\propto \rho_{de}$& $1+w$&$\xi$& $\rho_{dm}$ &$\rho_{de}$&${\bf d }$& Early time \\
& &  & & & &instability?\\
\hline \hline
& +& +& $\mp$ & +&+&Yes\\
 \hline
& +& --&+ & +&--&No\\
\hline
& --& --&+ & +&+&Yes\\
\hline
& --& +&$\mp$& +&--&No\\
\hline
\end{tabular}
\caption{\label{tab:rhoe} \em Scenario with $ Q\propto \mathcal  \rho_{de}$.
Stability criteria driven by the sign of ${\bf d }$ in Eq.~(\ref{eq:maldito_us}), for $|{\bf d }|>1$.
 The $\mp$ signs indicates that $\rho_{dm}$ is negative in the past for large positive couplings only.}
\end{center}
\end{table}

%

In the next subsection, we will confront in detail the model with data, sweeping over 
  values of the dark coupling strength $\xi<0$, although 
 models with $\xi>0$ are also viable as long as $\xi > -w$, as previously shown.

\subsection{Cosmological constraints}
\label{sec:Olga}

We explore here the current constraints on the dark energy-dark matter coupling $\xi$, considered constant, allowing for a non zero spatial curvature. The framework used is a cosmological model described by ten free parameters,
\begin{equation}
\theta=\left\{\omega_{b}, \omega_{dm}, \theta_{CMB}, \tau, \Omega_k, f_\nu, w, \xi, n_s, A_s \right\}~,
\end{equation}
where $\omega_{b}=\Omega_{b} h^2$ and $\omega_{dm}=\Omega_{dm} h^2$ are the
physical baryon and dark matter densities respectively,
  $\theta_{CMB}$\footnote{The $\theta_{CMB}$ parameter can be
  replaced by the $H_0$ parameter. However, using $\theta_{CMB}$ is
  better due to its smaller correlation with the remaining
  parameters.} is proportional to the ratio of the sound horizon to
the angular diameter distance, $\tau$ is the reionisation optical
depth, $\Omega_k$ is the spatial curvature, $f_\nu =
\Omega_\nu/\Omega_{dm}$ refers to  the neutrino fraction, 
  $n_s$ is the scalar spectral index and $A_s$ the scalar amplitude.
 The priors adopted on those parameters are given in Tab.~\ref{tab:priors}.

\begin{table}[tb]
\begin{center}
\begin{tabular}{|c||l|} \hline
 Parameter & Prior \\ \hline \hline 
$\omega_{b}$ &  0.005-0.1\\\hline 
$\omega_{dm}$ & 0.01-0.99\\\hline 
$\theta_{CMB}$ & 0.5-10\\\hline 
$\tau$ & 0.01-0.8\\\hline 
$\Omega_k$ &-0.1-0.1\\\hline 
$f_\nu$ & 0-0.3\\\hline 
$w$ & -1-0\\\hline 
$\xi$ & -2-0\\\hline 
$n_s$ & 0.5-1.5\\\hline 
ln($10^{10} A_s$) &2.7-4.0\\
\hline 
\end{tabular}
\caption{\label{tab:priors} \em Priors for the cosmological fit parameters considered in 
this work. All priors are uniform in the given intervals.}
\end{center}
\end{table}

We use the publicly available package
\texttt{cosmomc}~\cite{Lewis:2002ah}, modifying it in order to
include the coupling among the dark matter and dark energy
components, for the model considered in this section. 

A conservative compendium
of cosmological datasets is considered. First, what we call \emph{run 0} includes 
the WMAP 5-year data~\cite{Dunkley:2008ie,Komatsu:2008hk}, a prior on
the Hubble parameter of $72\pm8$~km/s/Mpc from the Hubble key
project (HST)~\cite{Freedman:2000cf}, the constraints coming
  from the latest compilation of supernovae~\cite{Kowalski:2008ez} and
  the $H(z)$ data\footnote{We thank R.~Jim\'enez and
    L.~Verde for suggesting the $H(z)$ data addition to the analysis.}  at $0<z<1.8$ from galaxy ages obtained by \cite{Simon:2004tf}.  We present a second data
analysis, that we call \emph{run I}, in which we add to \emph{run 0} the data on the
matter power spectrum (large scale structure data or LSS data) from the spectroscopic 
survey of Luminous Red Galaxies (LRGs) from the Sloan Digital Sky
Survey (SDSS) survey~\cite{Tegmark:2006az}. 

\begin{figure}[t]
\vspace{-0.1cm}
\begin{center}
\begin{tabular}{cl}
\hspace*{-0.75cm} 
\psfrag{c}[c][c]{\tiny{$m$}}
\includegraphics[width=8cm]{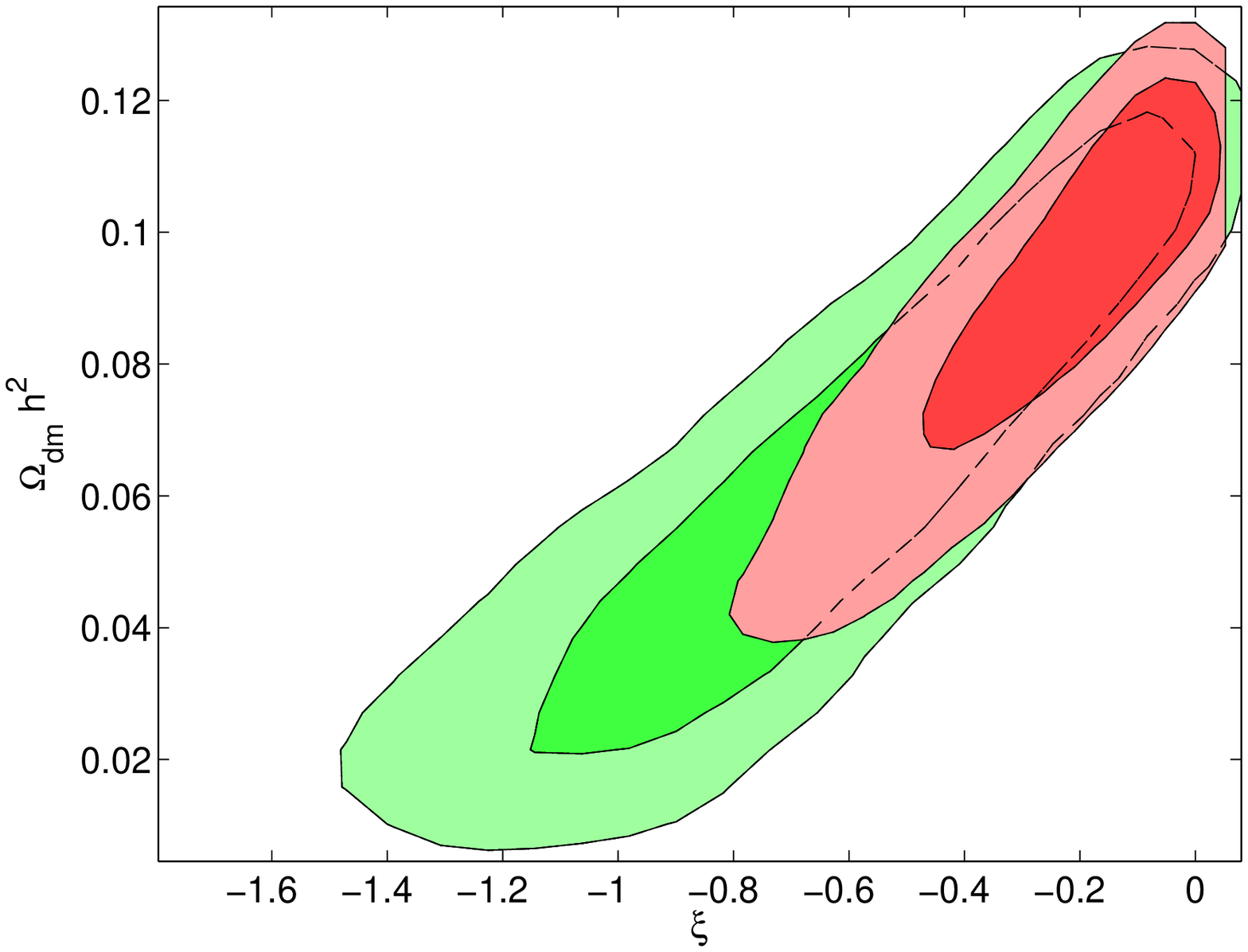} &
\includegraphics[width=8cm]{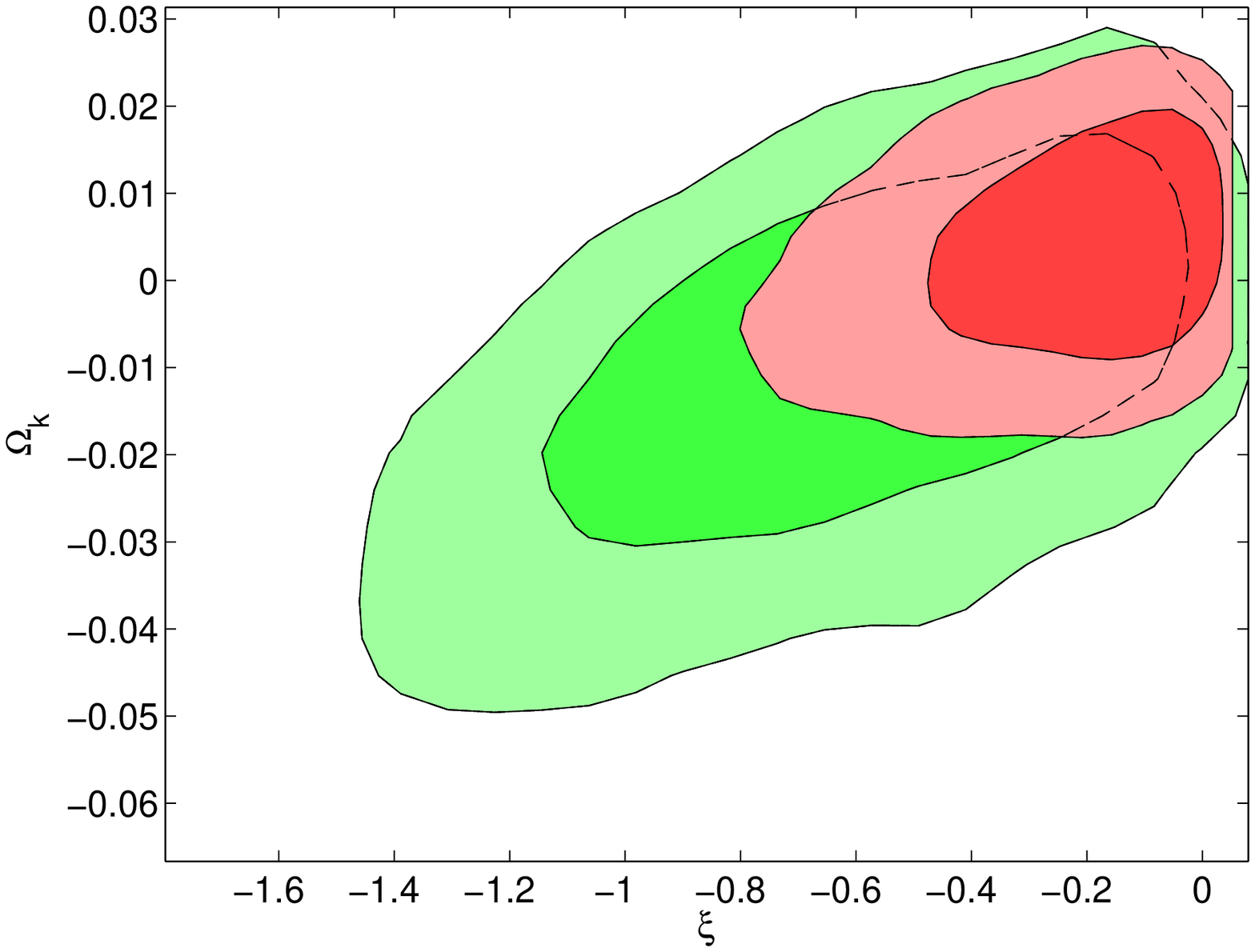} \\
\end{tabular}
\caption{\it Scenario with $Q\propto \rho_{de}$. Left (right) panel: 1$\sigma$ and 2$\sigma$ marginalized
  contours in the $\xi$--$\Omega_{dm} h^2$ ($\xi$--$\Omega_k$) plane. The
  largest, green contours show the current constraints from WMAP (5
  year data), HST, SN  and $H(z)$ data. The smallest, red contours
  show the current constraints from WMAP (5 year data), HST, SN,
  $H(z)$ and LSS data.}
\label{fig:fig0o}
\end{center}
\end{figure}

In summary, 
\begin{itemize}
\item \emph{run0}=WMAP(5yr)+HST+SN \textbf{+$H(z)$},
\item \emph{runI}=\emph{run0}+LSS.
\end{itemize} 
Figure~\ref{fig:fig0o} (left panel) illustrates the $1$ and $2\sigma$ marginalized contours in the $\xi$--$\Omega_{dm} h^2$ plane.
The results from the two runs described above are shown. We restrict
ourselves here to negative couplings and $w>-1$, which guarantees that 
  instability problems in the dark energy perturbation equations are avoided for all values of $\xi$. 
Notice that a huge degeneracy is present, being $\xi$ and $\Omega_{dm}
h^2$ positively correlated. The shape of the contours can be easily
understood. In a universe with a negative dark coupling $\xi$, the
matter content in the past is higher than in the standard
$\Lambda$CDM scenario due to an extra contribution proportional to
the dark energy component, see Eq.~(\ref{eq:adm}). Therefore, the amount of \emph{intrinsic} dark
matter needed -  that is, not including the contribution of dark energy through the coupling term -  should decrease as the dark coupling becomes more and
more negative and can be as small as 0.02, as indicated by \emph{run 0} results. The addition of LSS data to the analysis reduces considerably the
allowed parameter space. This is due to the enormous growth of clustering 
 for values of the coupling $\xi< -0.5$. For this range, the 
amplitude of the fluctuations increases, reaching values of $\sigma_{8}> 2$ and therefore providing a bad fit to LSS data\footnote{The authors of Ref.~\cite{Komatsu:2008hk} have reported a best fit value $\sigma_8 = 0.812 \pm 0.026$.}.

The right panel of Fig.~\ref{fig:fig0o} shows a positive correlation
between the coupling $\xi$ and the spatial curvature $\Omega_k$. High
precision CMB data indicates that currently the spatial curvature is a
subdominant contribution to the energy budget of the
universe\footnote{The authors of Ref.~\cite{Wang:2007mza} have found
  $\Omega^{(0)}_k =-0.002^{+0.041}_{-0.032}$ ($95\%$ CL limits)
  assuming a dynamical dark energy component.}, which implies
$\Omega_{de}^{(0)} + \Omega_{dm}^{(0)} + \Omega_{b}^{(0)} \simeq 1$. 
A non zero spatial curvature component implies instead
$\Omega_{de}^{(0)} +\Omega_{dm}^{(0)} + \Omega_{b}^{(0)} +  \Omega^{(0)}_k =1$. A negative coupling $\xi$ will increase the dark matter contribution and therefore a small negative curvature (closed universe) is needed to compensate the effect and describe well CMB data. The degeneracy between $\xi$ and $\Omega_k$ gets alleviated if one adds LSS data to the analysis.

\begin{figure}[t]
\vspace{-0.1cm}
\begin{center}
\begin{tabular}{cc}
\hspace*{-0.75cm} \psfrag{w}[c][c]{{$w$}}
\includegraphics[width=8cm]{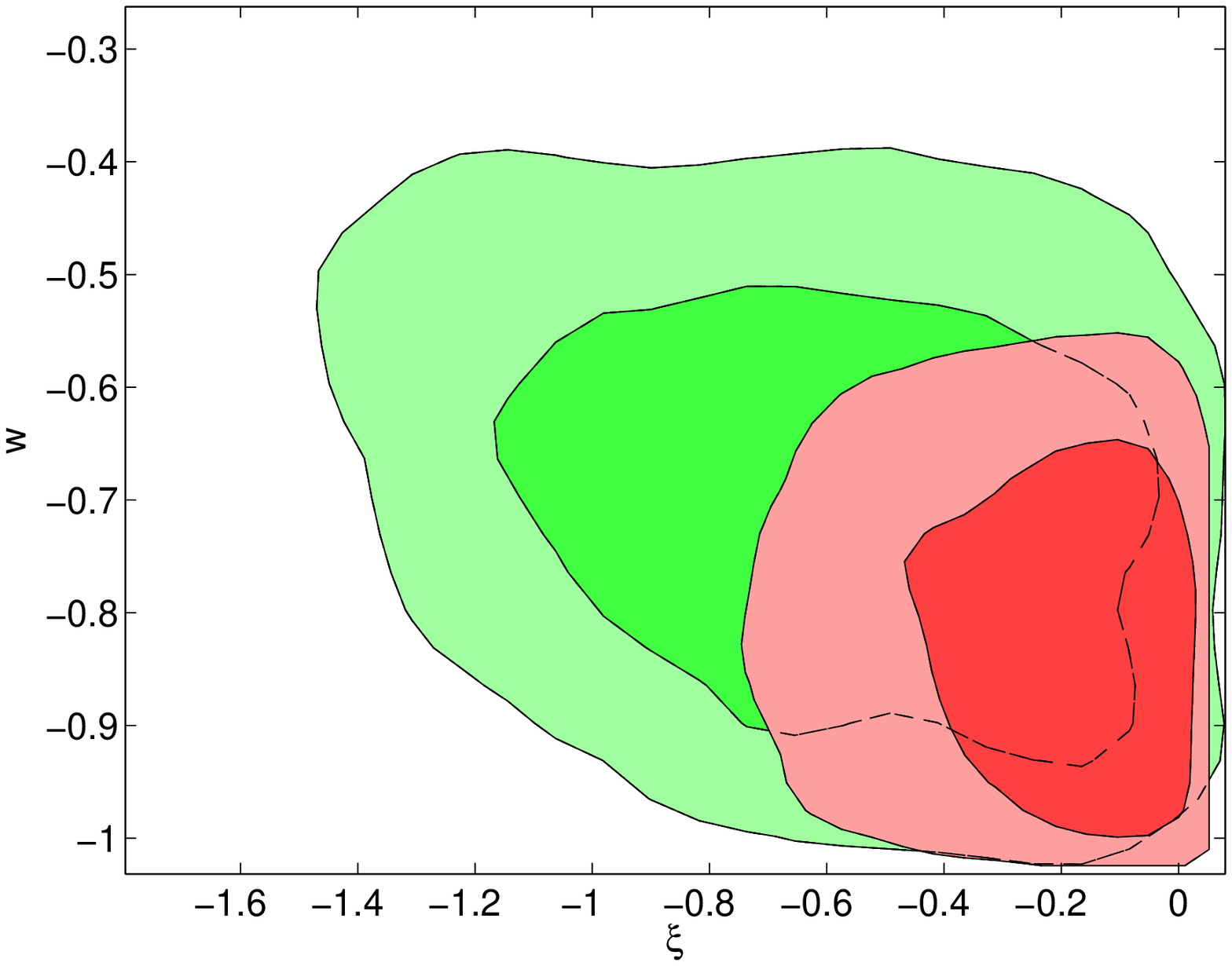} &
\includegraphics[width=8cm]{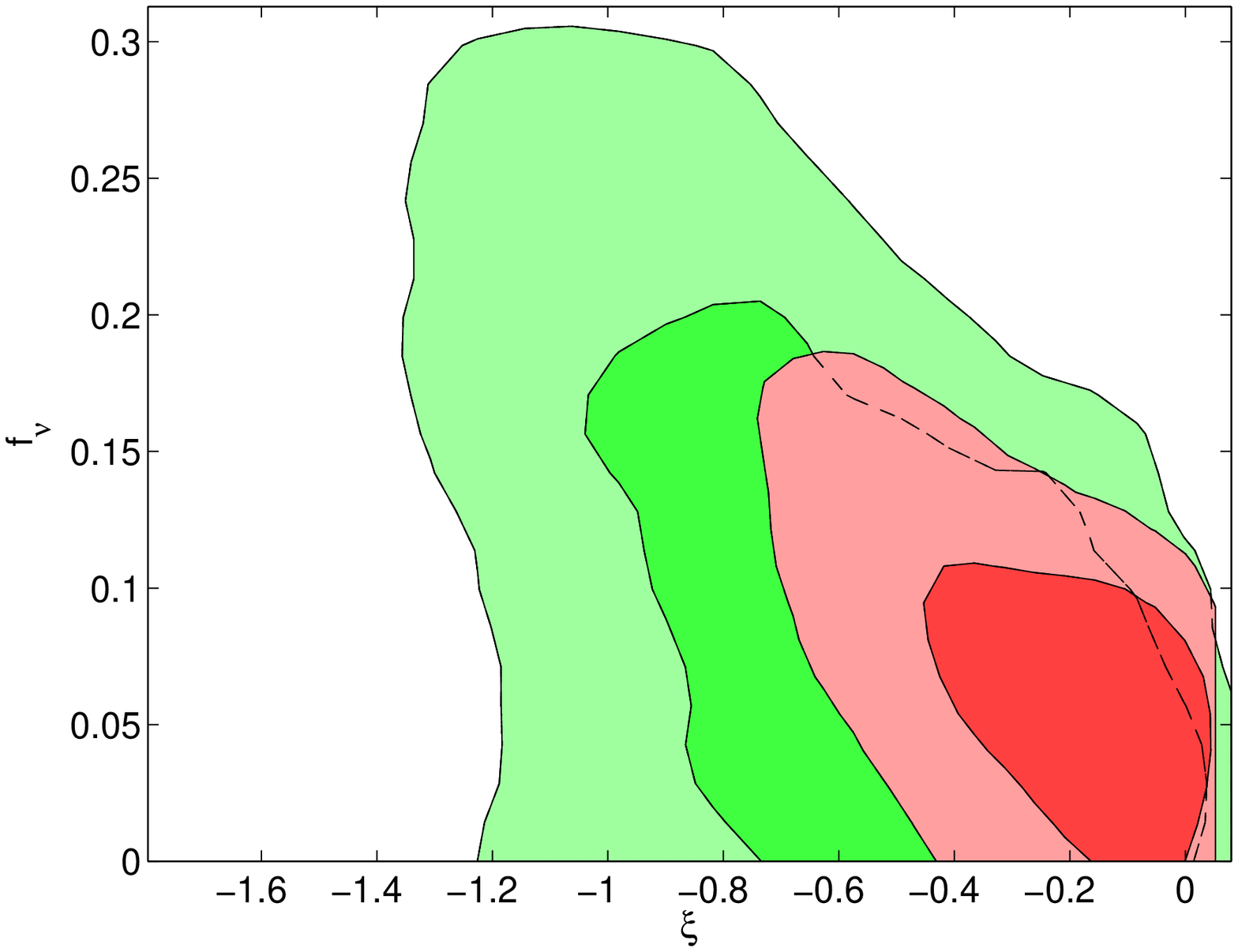} \\
\end{tabular}
\caption{\it Scenario with $Q\propto \rho_{de}$. Left (right) panel: 1$\sigma$ and 2$\sigma$ 
marginalized contours in the $\xi$--$w$ ($\xi$--$f_\nu$) plane. The largest, green contours show 
the current constraints from WMAP (5 year data), HST, SN and $H(z)$ data. The smallest, red contours 
show the current constraints from WMAP (5 year data), HST, SN, $H(z)$ and LSS data.}
\label{fig:fig1o}
\end{center}
\end{figure}

Figure~\ref{fig:fig1o} (left panel) depicts the constraints on the $\xi$--$w$ plane. 
We restrict ourselves here to $w>-1$ and $\xi <0$, a parameter region which ensures 
a negative doom factor, see Eq.~(\ref{eq:maldito_us}), and thus spans an instability--free 
region of scenarios to explore (see Sec.~\ref{sec:instab}).  Current data is unable to set 
strong constraints on the equation of state parameter $w$. 

\begin{figure}[h]
\vspace{-0.1cm}
\begin{center}
\begin{tabular}{c}
\hspace{-0.55cm} 
\includegraphics[width=10cm]{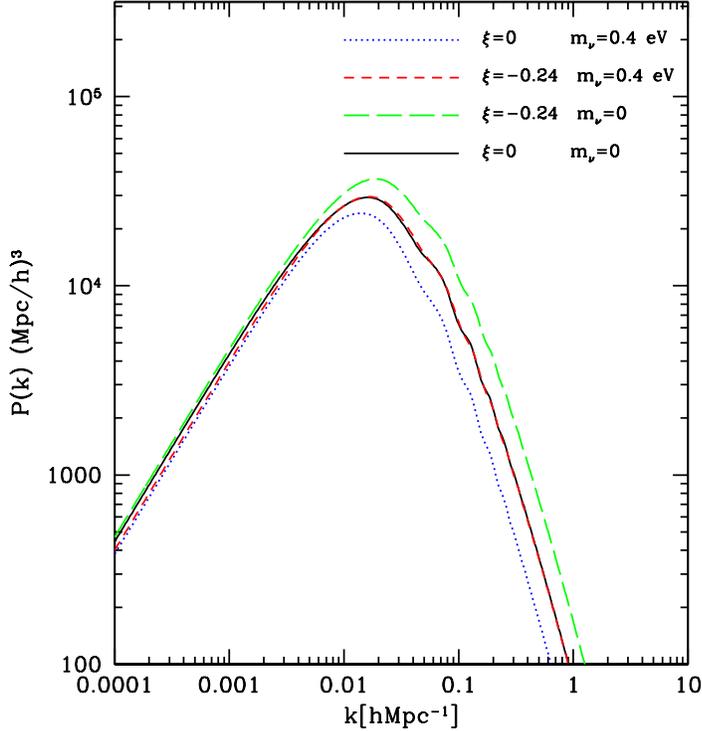} 
\end{tabular}
\caption{\it Scenario with $Q\propto \rho_{de}$. The black solid line depicts the matter power spectrum for the $\Lambda$CDM model with three massless neutrinos. The blue dotted line shows the mater power spectrum for a $\Lambda$CDM universe with three massive degenerate neutrinos with $\sum m_\nu =1.2$~eV. The green long--dashed curve refers to a universe with three massless neutrinos with a dark energy-dark matter coupling $\xi=-0.24$. The red short--dashed line illustrates a universe with $\sum m_\nu=1.2$~eV and a coupling $\xi=-0.24$.}
\label{fig:fig2o}
\end{center}
\end{figure}

The right panel of Fig.~\ref{fig:fig1o} shows next the correlation among the fraction of matter energy-density in the form of massive neutrinos $f_\nu$ and the dark coupling $\xi$. The relation between the neutrino fraction used here $f_\nu$ and the neutrino mass for $N_\nu$ degenerate neutrinos reads
\begin{equation}
f_\nu=\frac{\Omega_\nu h^2}{\Omega_{dm} h^2}=\frac{\sum m_\nu}{93.2 \textrm{eV}} \cdot \frac{1}{\Omega_{dm} h^2}=\frac{N_\nu m_\nu}{93.2 \textrm{eV}}\cdot \frac{1}{\Omega_{dm} h^2} ~.
\end{equation}
Neutrinos can indeed play a relevant role in large scale structure
formation and leave key signatures in several cosmological data sets,
see Ref.~\cite{Lesgourgues:2006nd} and references therein. More
specifically, the amount of primordial relativistic neutrinos changes
the epoch of the matter-radiation equality, leaving an imprint on both
CMB anisotropies (through the so-called Integrated Sachs-Wolfe effect)
and on structure formation, while non-relativistic neutrinos in the
recent Universe suppress the growth of matter density fluctuations and
galaxy clustering. This can be observed in Fig.~\ref{fig:fig2o}, where
the dotted curve depicts the matter power spectrum for three
degenerate massive neutrinos ($N_\nu=3$) with $m_\nu=
0.4$~eV. Notice that the matter power spectrum is reduced with respect
to the $m_\nu=0$ case, especially after the matter--radiation equality
era (imprinted in the power spectrum as a turnover). 


There is a strong and very well known degeneracy in the $\sum m_\nu-w$ plane, as first noticed in Ref.~\cite{Hannestad:2005gj}. Cosmological neutrino mass bounds become weaker if the dark energy equation of state is taken as a free parameter. If $w$ is allowed to vary, $\Omega_{dm}$ can take very high values, as required when $m_\nu$  is increased in order to have the same matter power spectrum. 
More recently, the authors of Ref.~\cite{LaVacca:2008mh} have pointed out that a higher neutrino mass is possible if dark matter and dark energy are coupled. 
Figure~\ref{fig:fig2o} shows the matter power spectrum in several
scenarios. The long dashed line refers to a universe with three
massless neutrinos and a coupling in the dark sector
$\xi=-0.24$. Notice that the matter power spectrum is enhanced with
respect to the $\xi=0$ scenario, due to the higher matter energy
density in models with a non-negligible coupling. However, the power
enhancement effect induced by the presence of a coupling can be
compensated by adding massive neutrinos in the game. Those neutrinos
will reduce the power spectrum, see the short dashed curve in
Fig.~\ref{fig:fig2o}, which is indistinguishable from the matter power
spectrum in a $\Lambda$CDM universe. This $m_\nu-\xi$ degeneracy is
shown in Fig.~\ref{fig:fig1o} (right panel): a neutrino mass of $\sum m_\nu\sim 1.5$~eV is allowed for couplings $\xi>-0.6$ at the 2$\sigma$ level.


\section{Comparison with the literature: $Q\propto\rho_{dm}$}
\label{sec:Val}
Early time non-adiabatic instabilities were pointed out in Ref.~\cite{Valiviita:2008iv}, in which 
several coupled models were considered. In particular they studied an interaction rate of
the form: $Q = \xi \mathcal{H} \rho_{dm}$, with constant $w$ and $\xi$, concluding that 
it was unstable for constant $w$ even for very small values of the coupling. We will readdress 
that model here, to clarify if and when it is subject to non-adiabatic instabilities and explore 
some supplementary aspects.

In this scenario, Eqs.~(\ref{eq:conservDM}) and (\ref{eq:conservDE}) lead to  the following 
background equations:
\begin{eqnarray}
  \label{eq:EOMmV}
  \dot\rho_{dm}+ 3\mathcal{H}\rho_{dm} &=&\xi\mathcal{H} \rho_{dm}\,,\\
\label{eq:EOMeV}
 \dot\rho_{de}+ 3 \mathcal{H}\rho_{de}(1+ w)&=& -\xi \mathcal{H} \rho_{dm}\,. 
\end{eqnarray}
These two fluids exhibit thus effective equations of state given by
\begin{eqnarray}
  &w_{dm}^{eff}= -\frac {\xi}{3 }\label{weff_dmV}\,,\\
  &w_{de}^{eff}= w+\frac {\xi}{3 }\frac{\rho_{dm}}{\rho_{de} }\,.
  \label{weff_de_V}
\end{eqnarray}
Notice that now the effective dark energy equation of state is redshift-dependent, 
and consequently, for constant $w$ and $\xi$,
\begin{eqnarray}
  \rho_{dm}&=& \rho_{dm}^{(0)} a^{-3+\xi}\,,\label{eq:admV}\\
  \rho_{de}&=& \rho_{de}^{(0)}a^{-3(1+w)}+ \rho_{dm}^{(0)}\frac{\xi}{\xi+3w}
      \left(a^{-3(1+w)}- a^{-3+\xi}\right)\,.\label{eq:adeV}
\end{eqnarray}
\begin{figure}[t]
\begin{center}
\includegraphics[height=.4\textheight]{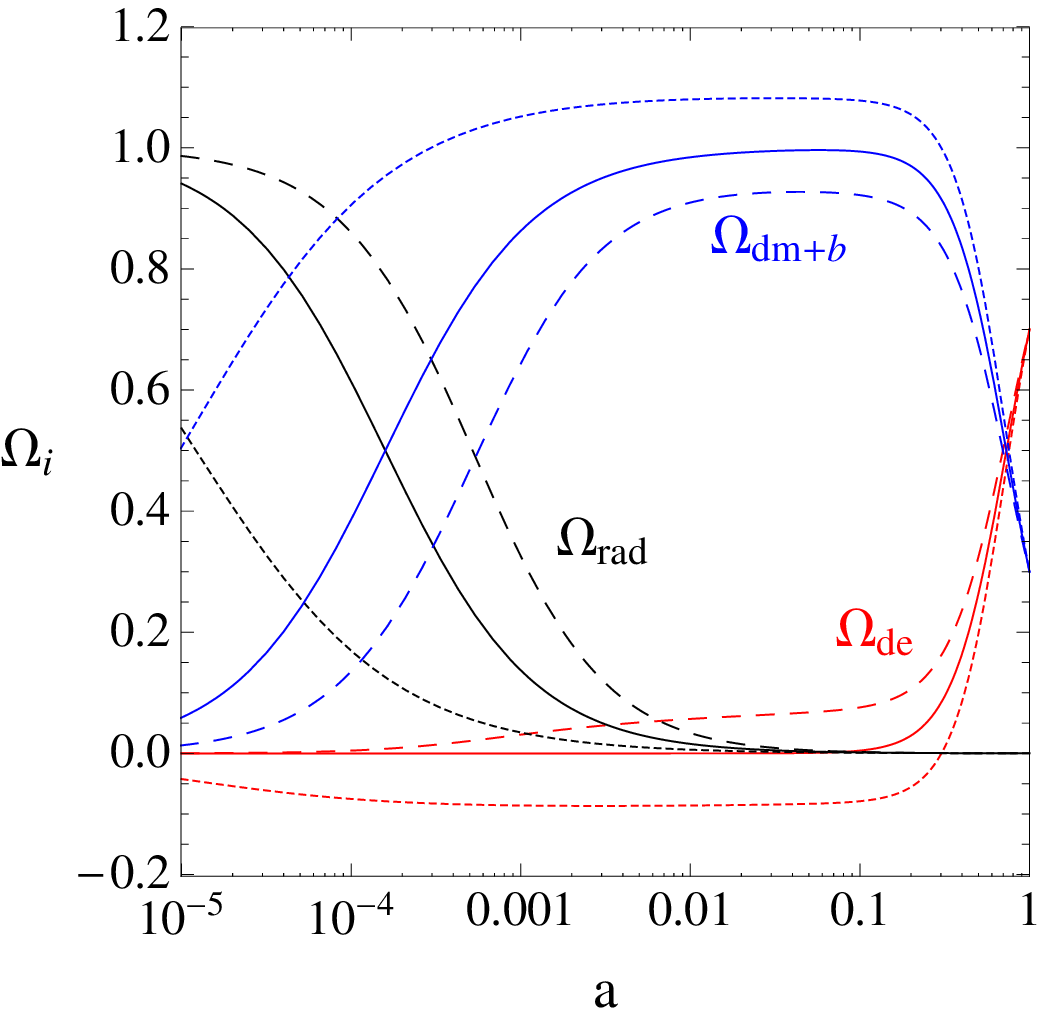}
\caption{\it
Scenario with $Q \propto \rho_{dm}$. Relative energy densities of dark matter + baryon 
$\Omega_{dm+b}$ (blue), radiation $\Omega_{rad}$ (black) and dark energy $\Omega_{de}$ 
(red), as a function of the scale factor $a$, for $w=-0.9$. Three values of the coupling are illustrated: 
$\xi= 0$ (solid curve), $0.25$ (long dashed curve) and $-0.25$ (short dashed curve).}
  \label{fig:fishesplotsV}
\end{center}
\end{figure}
The authors of Ref.~\cite{Valiviita:2008iv} chose to  restrict their analysis to constant negative 
$\xi$ and $1+w >0$. For such parameter values, in their model the dark energy density $\rho_{de}$ 
is negative  for any $w<0$~\footnote{And thus not only within 
the $-2/3>w>-1$ region explored in Ref.~\cite{Valiviita:2008iv}. Notice that in our convention 
$\xi=-\alpha$, with $\alpha$ defined in that work.}. This is illustrated in Fig.~(\ref{fig:fishesplotsV}), 
where the short dashed curve corresponds to $\xi=-0.25 $ values and $\Omega_{de}$ is seen 
to be negative since the universe birth until late times, while $\Omega_{dm+b}$ is always positive. 
This is in stark contrast with the behaviour in the scenario with $Q\propto \rho_{de}$ discussed in Sect.~\ref{ourmodel}, for which all energy densities remain positively defined all through the universe 
history. Figure~(\ref{fig:fishesplotsV})  exemplifies numerically  as well how positive (negative) values 
of $Q$ ameliorate (worsen) the coincidence problem, although again the effect is quantitatively 
unimportant for the phenomenologically allowed values of $\xi$. 

Moreover, the authors of Ref~\cite{Valiviita:2008iv} found that the instabilities appear no matter
how weak the coupling is. Before turning to the analysis of the doom factor and instabilities for 
this model, let us explore some further background-dependent properties.

\subsection{Reconstructing $\tilde w(z)$}
\label{sec:wzV}
\begin{figure}[h!]
 \begin{center}
\includegraphics[height=.35\textheight]{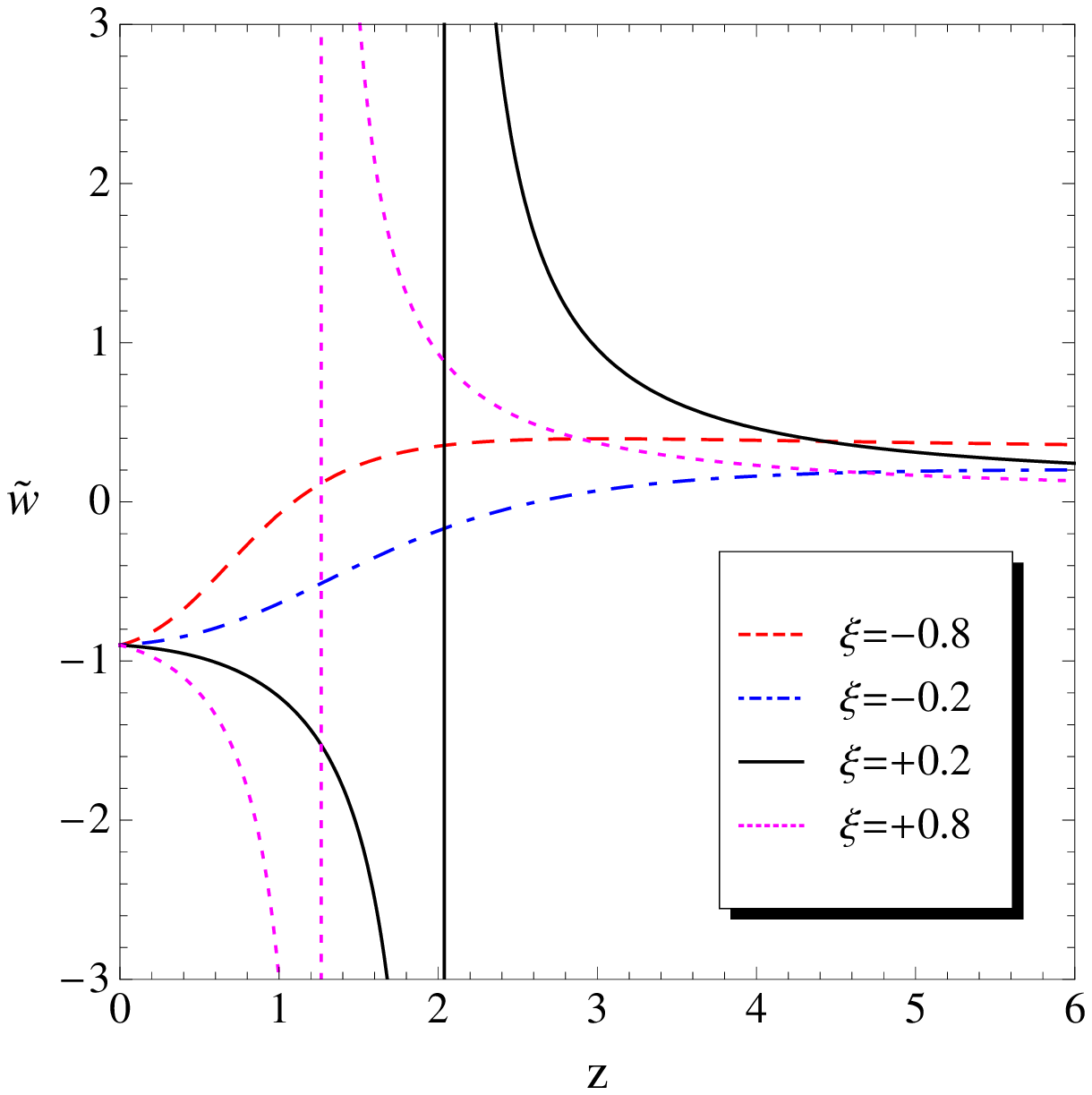}
\caption{\it
Scenario with $ Q\propto \mathcal \rho_{dm}$. Reconstructed $\tilde w(z)$ as function of $z$. 
We have considered $w= -0.9$. The black (solid) and magenta (short dashed) curves depict the 
$\tilde w(z)$ behaviour for $\xi=0.2$ and $\xi= 0.8$. The blue (long-short dashed) and the red 
(long-dashed) curves denote the $\tilde w(z)$ behaviour for $\xi=-0.2$ and $\xi=-0.8$.
Notice that for positive values of $\xi$ we recover the divergent phantom-crossing behavior 
appearing in scalar-tensor theories.}
  \label{tab:wzreconsV}
\end{center} 
\end{figure}

For this scenario, the reconstructed equation of state $\tilde w(z)$, if obtained from data mainly 
sensitive to the fluid background  and analyzed assuming no dark coupling, will diverge for positive 
couplings and redshifts around $z\sim\mathcal O (1-10) $. It will show a \emph{phantom crossing} 
behavior typical of scalar-tensor dark energy models, see Fig.~\ref{tab:wzreconsV}. Indeed using 
Eq.~(\ref{eq:wH}), we get a somewhat more complicated expression than for the $Q\propto 
\rho_{de}$ scenario: 
 \begin{equation}
   \label{eq:wzdeltaV}
\tilde w(z)=w\frac{\xi- (1+z)^{3w+\xi}  \left[( \xi+3w)r +\xi\right]}
         {-3w +(1+z)^{\xi}(3w+\xi)-(1+z)^{3w+\xi} \left[( \xi+3w)r +\xi\right]} \, 
\end{equation}
with $r= \Omega^{(0)}_{de}/\Omega^{(0)}_{dm}$.
 At small redshifts it reduces to:
\begin{equation}
   \label{eq:wznear-1}
\tilde w(z)\sim w\left(1 +\frac{\xi}{r} z \right)\, ,
\end{equation}
while at large redshifts the reconstructed $\tilde w(z)$ in Eq.~(\ref{eq:wzdeltaV}) obeys
\begin{eqnarray}
 \tilde w(z)\simeq \frac{-\xi}{3}\frac{1}{1-z^{\xi}} \rightarrow & 0 &
 \quad \mbox{ for }\quad  \xi>0\,, \\
 \rightarrow & \displaystyle -\frac{\xi}{3} & \quad \mbox{ for }\quad  \xi<0\,,
\end{eqnarray}
assuming $|\xi|< 1$.

\subsection{Linear perturbation theory}
\label{sec:linV}

The propagation of dark energy pressure waves is again driven by Eq.~(\ref{eq:dpcs}), 
with the doom factor Eq.~(\ref{eq:maldito}) reading now:
\begin{equation}
\label{eq:maldito_vali}
{\bf d }= \frac{\xi}{3(1+w)}\,\frac{\rho_{dm}}{\rho_{de}}.
\end{equation}
This expression should be compared to the analogous one when $Q\propto \rho_{de}$, see 
Eq.(\ref{eq:maldito_us}): they differ in the $\rho_{dm}/\rho_{de}$ factor in the dark coupling term, 
 which can be large at early times when $\rho_{de}$ may become negligible. 
At early times, the doom factor will always thus be large even for tiny values of $\xi$, 
as the ratio $\rho_{dm}/\rho_{de}$ is very large then.
 This ratio is the major difference between the scenarios with $Q\propto \rho_{dm}$ 
(or $Q\propto \rho_{dm}+\rho_{de}$ and alike) and those in which $Q\propto \rho_{de}$. 
As the true expansion factor is ${\bf d}$ instead of $\xi$, this fact also explains why in
the present scenario the phenomenological analysis only allows for $\xi$ values much 
smaller in magnitude than those permitted for $Q\propto \rho_{de}$. 

The evolution of the perturbations, in the synchronous comoving gauge, is described in this scenario 
by Eqs.~(\ref{eq:deltambis}), (\ref{eq:deltaebis}) and (\ref{eq:thetaebis}) with 
\begin{eqnarray}
&\,&\delta\left[{\bf d}\right]=\,\frac{\xi}{3(1+w)}\,
\frac{\rho_{dm}}{\rho_{de}}(\delta_{dm}-\delta_{de})\,,\label{eq:deltad_vali}\\
&\,&\delta\left[\frac{\rho_{de}}{\rho_{dm}}\,{\bf d}\right]= \,0\,.\label{eq:deltak_vali}
\end{eqnarray}

\subsection{Early time (in)stability}
\label{sec:instabV}

The strong coupling regime is defined by $|{\bf d}|>1$, as explained in Sect.~\ref{strongcoupling}.
In this regime and for large scales, $\mathcal H/k \gg 1$, the dark energy perturbations are 
described by Eqs.~(\ref{eq:deltaxi}) and (\ref{eq:thetaxi}), for which in this case 
Eqs.~(\ref{eq:maldito_vali}), (\ref{eq:deltad_vali}) and
(\ref{eq:deltak_vali}) above apply. It leads, for constant $w\ne-1$,
to a $\delta_{de}$ and $\delta_{de}'$ contribution to the early time growth at large scales, described by
\begin{eqnarray}
\label{growthvali}
\delta_{de}''  & \simeq &   3(\hat c_{s\,de}^2+1)\,{\bf d } \,\, 
  \Big[\, \left(3 \,\frac{\hat c_{s\,de}^2-w}{\hat c_{s\,de}^2+1}\,-\, (1+w)\,{\bf d }\right)\, 
  \frac{\delta_{de}}{a^2}+ \frac{\delta_{de}'}{a}\Big]+ ...
\end{eqnarray}
where dots account for all other terms, which will be subdominant whenever the two terms above induce by themselves an exponential growth.

Once again, for ${\bf d }>0$ the $\delta_{de}'$ term signals an antidamping -growing- regime,  
which may induce instabilities, whenever ${\bf d}>1$, when combined  with either a positive or a 
negligible negative $\delta_{de}$ coefficient. Interestingly, Eq.~(\ref{growthvali}) shows that 
the latter is determined in this case by a competition between the two factors within brackets.  
We have verified numerically that, for values of $|\xi|<1$ and $w$ around
-1,  both terms are of the same order and strongly cancel. 
For positive ${\bf d }$, all cases of Sec.~\ref{sec:instabV} with $A_e>0$ or $<0$ and $B_e>0$ 
develop then a strong instability at early time, see upper panels in Fig.~\ref{fig:strongV}.

Turning back to the characteristics of the doom factor in this model, Eq.~(\ref{eq:maldito_vali}), 
it is to be noted that  the ratio $\rho_{dm} / \rho_{de}$ not only enhances in the past the magnitude
of ${\bf d }$ and thus the onset of the strong coupling regime even for very small $\xi$ values, 
but it also influences its sign. Indeed, recall that $\rho_{de}$ is negative in the past for negative 
$\xi$, as can be seen from Eq.~(\ref{eq:adeV}), with an obvious impact on the sign of {\bf d}.

The identification of the ${\bf d }$ factor allows thus to predict the range of parameters 
for which this model will be {\it stable}, even with constant $w$. We have summarized the analysis 
of the early time (in)stability criteria for a model with $Q\propto \mathcal \rho_{dm}$ in Tab.~\ref{tab:rhom}.  
\begin{table}[bt]
\begin{center}
\begin{tabular}{|c||c|c|c|c|c|c|} \hline
 Model: $ Q\propto \mathcal \rho_{dm}$ & $1+w$&$\xi$& $\rho_{dm}$ &$\rho_{de}$&$d$&Early time \\
 & &  & & & &instability?\\
\hline \hline
& +& -- & + & $\mp$ &+&Yes\\
 \hline
& +& +&+ & +&+&Yes\\
\hline
& --& +&+ & +&--&No\\
\hline
& --& --&+& $\mp$ &--&No\\
\hline
\end{tabular}
\caption{\label{tab:rhom} \em Scenario with 
$Q \propto \rho_{dm}$. Stability criteria driven by the sign of ${\bf d}$ in Eq.~(\ref{eq:maldito_vali}),
whenever $|d|>1$. The $\mp$ signs indicates that $\rho_{de}$ is negative in the past 
 for negative couplings independently of their value.}
\end{center}
\end{table}
It is illustrative and amusing to further analyze these patterns with some examples. Figure~\ref{fig:strongV} 
shows two cases of early time instabilities in the upper panels, and two examples of early time 
stability in the lower panels:
\begin{figure}[!ht]
\begin{center}
 \begin{tabular}{c c} 
  \includegraphics[width=8cm]{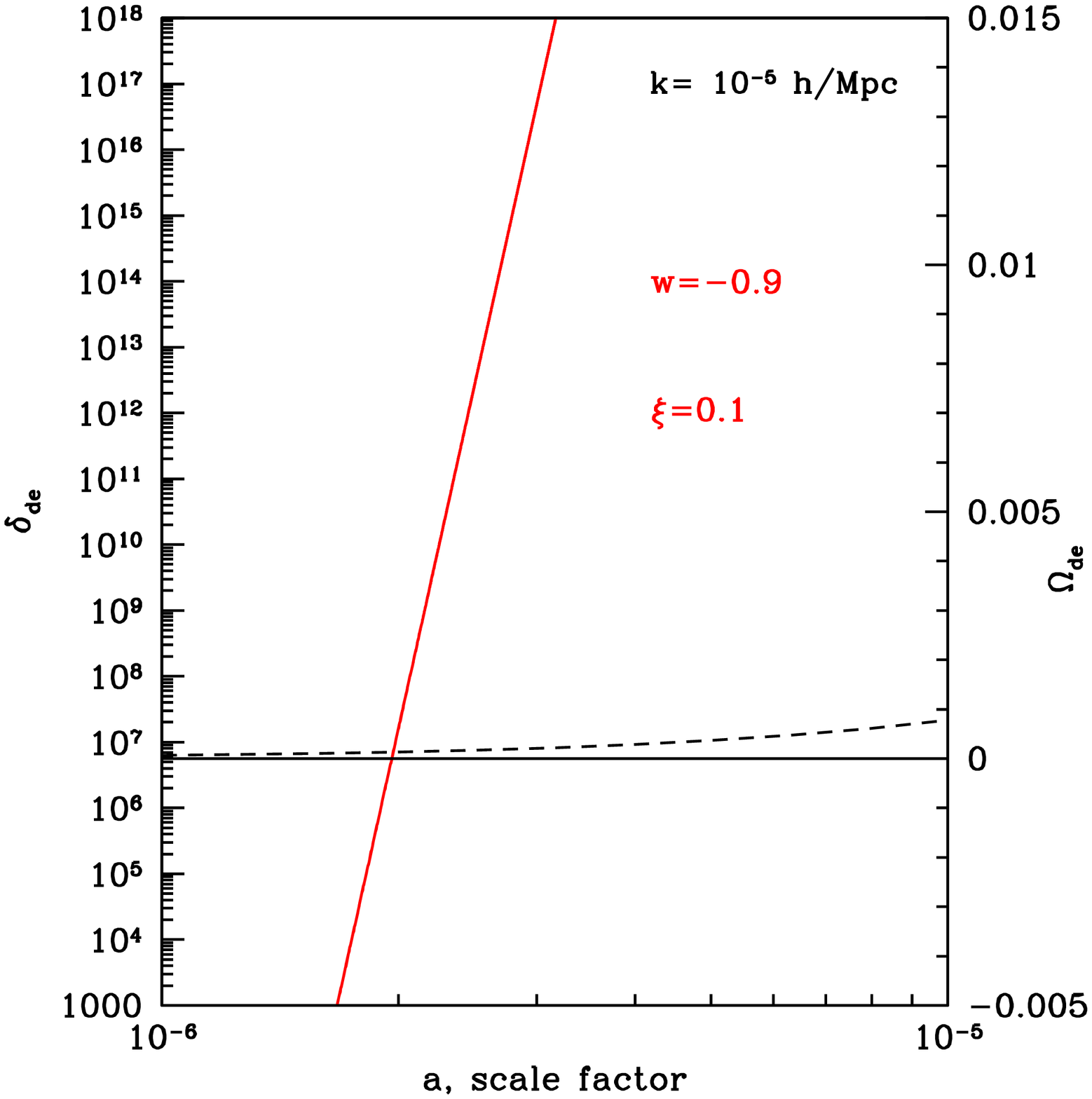}&
  \includegraphics[width=8cm]{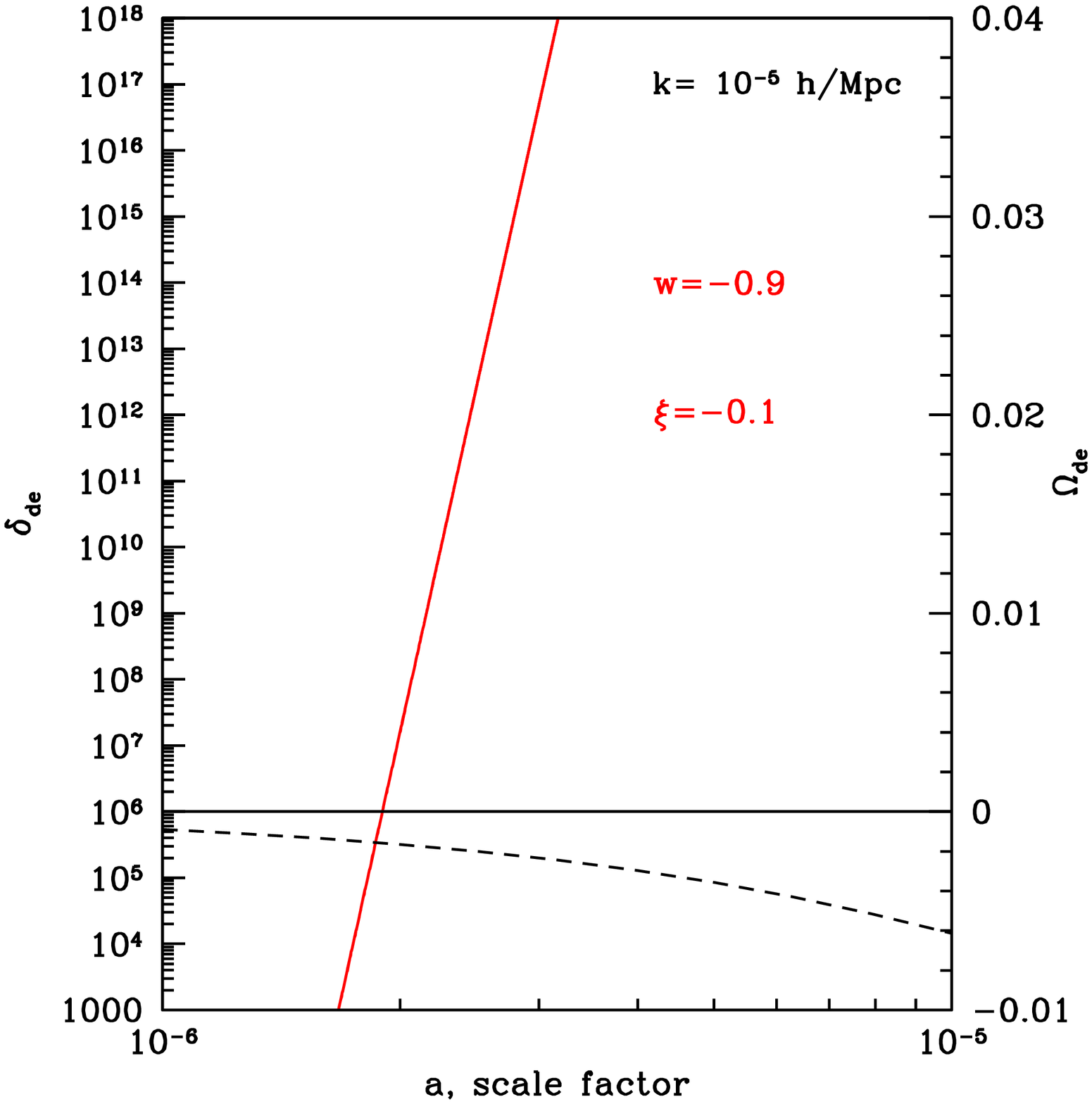}\\
  \includegraphics[width=8cm]{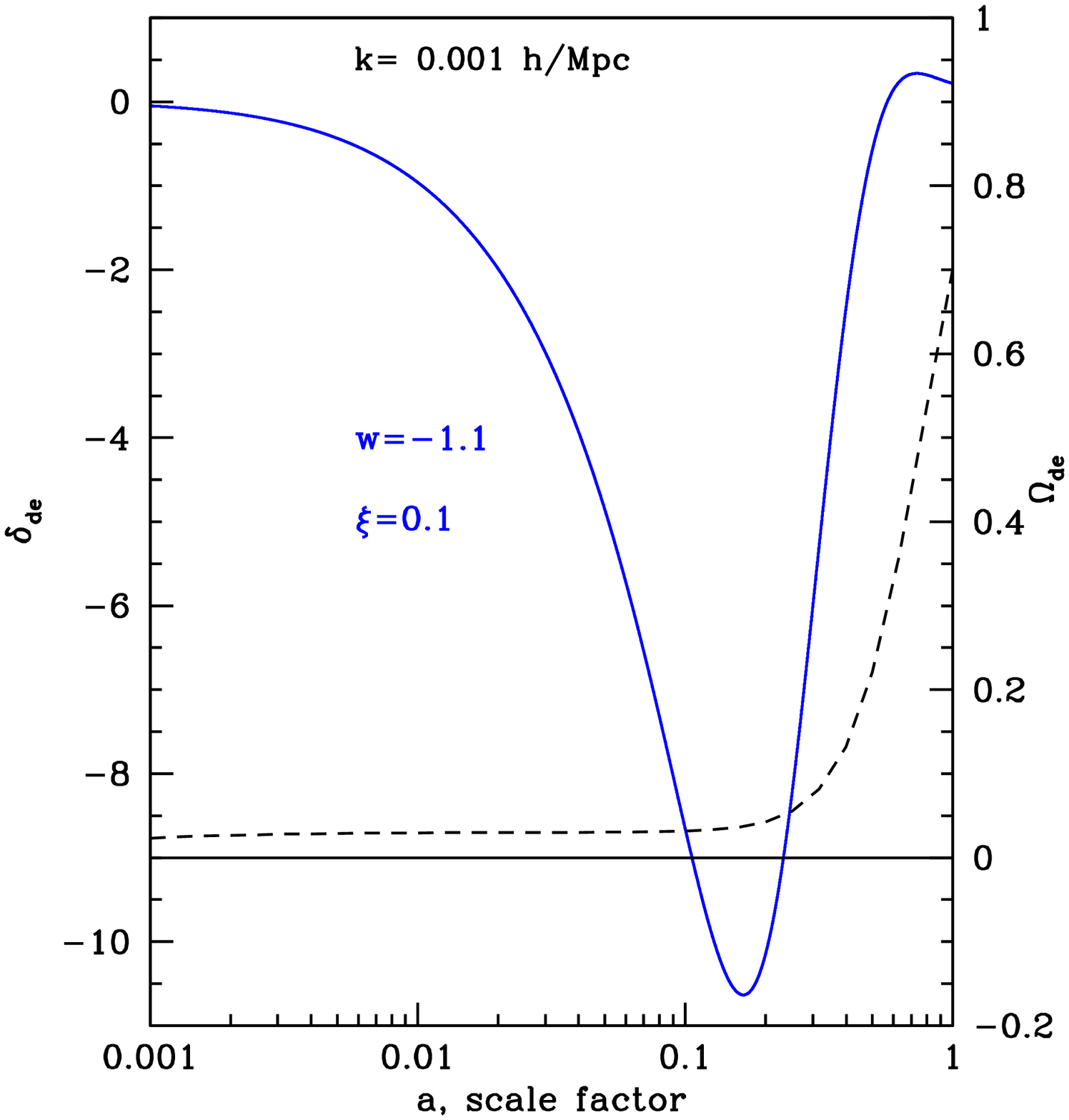}&
  \includegraphics[width=8cm]{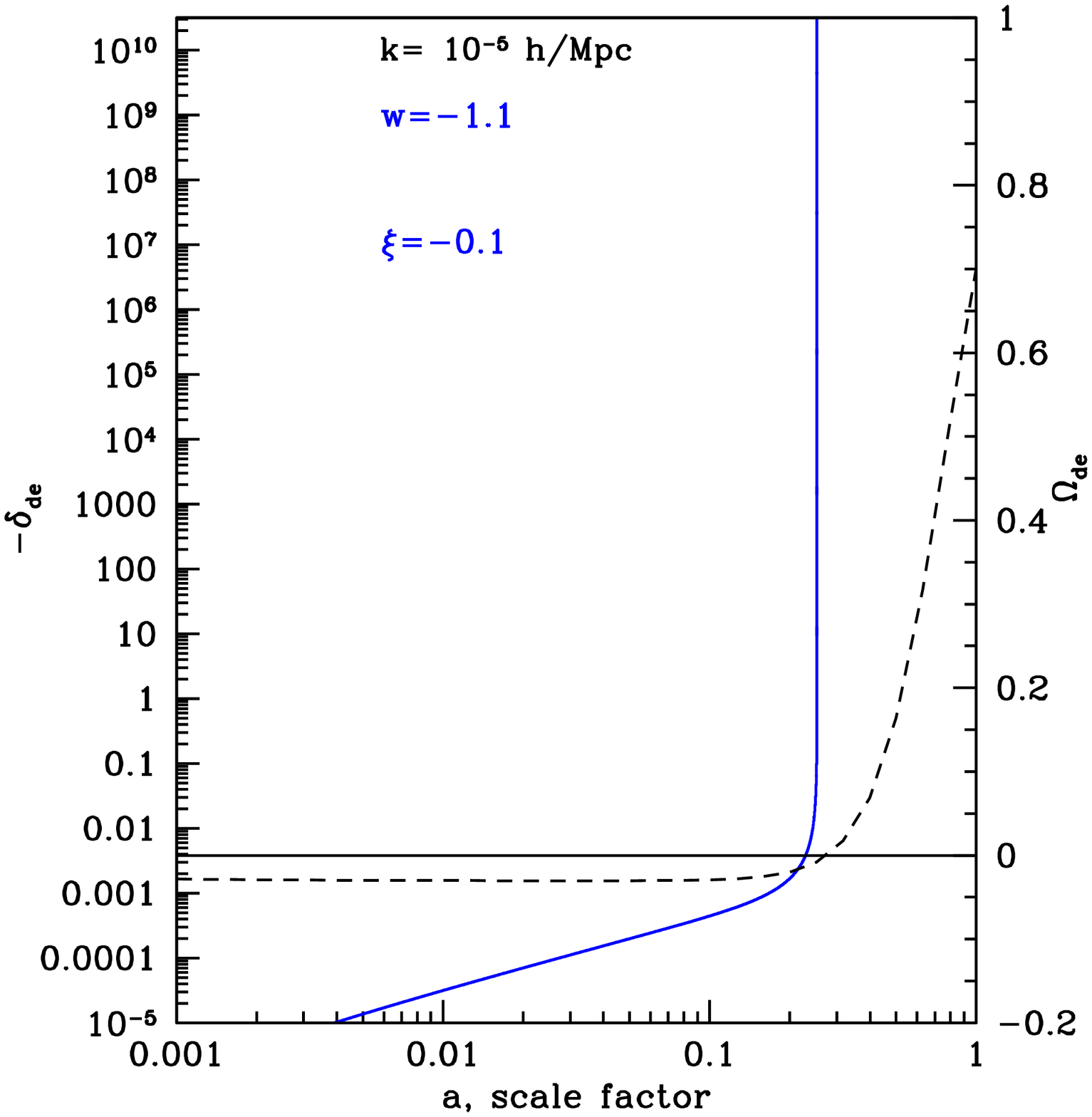}\\
 \end{tabular}
\caption{\it $Q \propto \rho_{dm}$. Upper panels: Evolution for the $\delta_{de}$
perturbation vs the scale factor at scales $k=10^{-5}$ h/Mpc for $w=-0.9$ and 
$\xi=\pm 0.1$. Early time instability is present in both cases, as 
predicted from our analytical study.  The black dashed curve shows the positive 
(negative) value of the dark energy density $\Omega_{de} (a)$ (right scale). 
The lower right (left) panel shows the $\delta_{de}$ perturbation vs the scale 
factor at scales $k=10^{-5}$ ($k=0.001$) h/Mpc for $w=-1.1$. For both positive 
and negative couplings the
 model is free of early-time instabilities. However, for negative couplings, there exists 
a late-time instability (see right panel), caused by a change in the sign of the dark 
energy density $\Omega_{de} (a)$, shown by the black dashed curve (right scale). 
}
  \label{fig:strongV}
\end{center} 
\end{figure}
\begin{itemize}
\item The two upper graphics both assume $1+w>0$, while the dark coupling strength $\xi$ 
has opposite sign in the right and left panels. Nevertheless, both scenarios  are unstable 
from their birth, as the sign of $\rho_{de}$ is also opposite as can be seen from Eq.~(\ref{eq:adeV}) 
and depicted as a discontinuous line in the figures, resulting in a doom factor ${\bf d }$  
- Eq.~(\ref{eq:maldito_vali}) - which is positive for both scenarios. 
\item The two lower panels analyze the analogous cases although with $1+w<0$ instead. 
Again, at early times both models have opposite sign as well for the dark energy density 
$\rho_{de}$, combining into a safe negative value for ${\bf d }$.
\item A late time instability is seen to appear, though, for the model in the lower right panel. 
Indeed, this is well understood as for this model $\rho_{de}$  is seen to change sign during cosmic 
evolution, and in consequence precisely at that temporal point the model finds its {\it doomsday}, 
as ${\bf d }$ becomes positive.
\end{itemize}
This latter example pinpoints that our (in)stability criteria on the growth of dark energy perturbations 
also applies to late evolution, if handled with care.

Finally, as pointed out in Ref.~\cite{Valiviita:2008iv}, a possible way to avoid the early time instability 
of their models is to relax the assumption of a constant equation of state of the dark energy 
component and allow instead for a dynamical behavior, i.e, for a time-varying $w(a)$, even when 
the dark coupling is turned off. For instance, the CPL~\cite{Chevallier:2000qy,Linder:2002et} 
parameterization $w(a)=w_{0}+w_a (1-a)$ can alleviate the instability problem because of 
its smaller ratio $\rho_{dm}/\rho_{de}$ in the past. From our perspective, this may allow those 
models to avoid in the past the strong coupling regime as we defined it, trading it for a 
\emph{softly coupled model}, that is, one in which the evolution is dominated by the 
coupling-independent terms and thus alike to the stable uncoupled regime discussed in 
Sec.~\ref{sec:uncoupl}. A detailed perturbation analysis with different dark energy 
parameterizations and/or time dependent couplings will be presented elsewhere \cite{newUS}.

Note as well that models with $Q\propto (\alpha \,\rho_{dm}+\beta \,\rho_{de})$, 
with $\alpha$ and $\beta$ arbitrary constants and $\alpha\ne 0$~\cite{CalderaCabral:2008bx} 
will  also be subject to reinforced behaviour at early times, as in these scenarios the dark energy 
perturbations will generically present as well a dependence of the coupling terms on $1/\rho_{de}$. 
The corresponding stability analysis and criteria will parallel that
of the scenario analyzed in this section.

\section{Conclusions}

We have considered the possibility that dark energy -the engine for accelerated expansion of the universe- is not pure vacuum energy but has some dynamical nature. In these conditions, dark matter and dark energy may evolve independently or be coupled. We have allowed for such a coupling, the {\it dark coupling}, and explored the conditions for a viable universe and its phenomenological signals. The generic evolution of both the cosmic background and the coupled dark matter-dark energy density perturbations within the linear regime has been developed.
 
In particular and  without referring to any particular model, we have identified the origin of 
non-adiabatic large scale instabilities at early times, which affect many coupled models with constant 
$w$, where $w$ would be the dark energy equation of state in the absence of dark-coupling.
The instability is related to the presence of the coupling terms in the propagation of 
dark energy pressure waves within the dark matter
background~\cite{Valiviita:2008iv}. We have shown that the size and
sign of the dark-coupling terms is essential, identifying the combination of parameters which characterize the (un)stable regimes: the doom 
factor ${\bf d}$ given by the ratio
\begin{equation}
{\bf d }\,= \frac{Q}{3{\mathcal H } \rho_{de} \, (1+w)}\,, 
\end{equation}
where $Q$ encodes the interaction rate between the two dark sectors. 
We showed that when ${\bf d }$ is positive and sizeable, ${\bf d }>1$, the dark-coupling 
dependent terms may dominate the evolution of dark energy perturbations, which will then enter 
a runaway, unstable, growth regime. We have thus established a general condition necessary 
to obtain models free from  non-adiabatic instabilities at large scales. Although for the sake 
of precision the analysis has concentrated on early time instabilities,  it also sheds light 
on late time non-adiabatic ones.

The insight provided by the analysis allows to predict, for any given model, which range 
of parameters  may result in an {\it a priori} stable universe.  As an illustration, we have 
analyzed in detail  the simple scenario in which $Q$ is proportional to the evolving dark energy 
density, $Q=\xi {\mathcal H} \rho_{de}$.  In addition to the analytical study, we have used 
the publicly available codes \texttt{CAMB} and \texttt{cosmomc}, modifying them to account for 
the interaction between the two dark sectors, and considering WMAP-5 year data, HST data, 
supernova data, $H(z)$ data and large scale data structure from the SDSS survey.

Models of the class discussed in the previous paragraph, with  negative dark-coupling $\xi<0$ and positive $(1+w)$, 
besides being stable as ${\bf d }<0$, give the best agreement with data on large scale 
structure formation~\footnote{Although models with $\xi<0$ worsen the coincidence 
problem, while for $\xi>0$ it gets alleviated, these effects are quantitatively minor in viable
models. }.  For them, both $w$ and $\xi$ are not very constrained from data, and large values 
for both parameters, near -0.5, are easily allowed. Furthermore, 
$\xi$ turns out to be positively correlated with both  $\Omega_{dm} h^2$ and the curvature 
$\Omega_k$. The results show as well the neutrino mass  - dark coupling degeneracy: i.e. a
neutrino mass of $\sum m_\nu \sim 1.5$~eV is allowed for couplings $\xi>-0.6$,  at the 
$2\sigma$ level. In resume,  the scenario satisfies all current constraints from WMAP, HST, 
SN, LSS and $H(z)$ data and is free from instabilities, including early ones. Future
direct measurements of $H(z)$, as those provided by BAO surveys, might be crucially 
important to improve the current bounds on the dark coupling $\xi$. A detailed analysis 
will be presented elsewhere~\cite{newraul}.

Within the same class of models, positive values of $\xi<-w$ are not excluded (with $1+w<0$ 
to ensure ${\bf d }<0$), and enjoy the smoking-gun signal of inducing an apparently divergent reconstructed equation 
of state, at  redshifts  $z-{\mathcal O}(1-10)$, when extracted  from background-dominated 
data and analyzed assuming no-coupling.  It is remarkable that, in contrast, no such divergent 
``crossing of the phantom divide" occurs for any $\xi<0$.

Our results have also clarified the origin of early time non-adiabatic instabilities found in 
previous models in the literature in which $Q\propto \rho_{dm}$ and $Q\propto \rho_{dm}+
\rho_{de}$, and we have analyzed the former class in detail, up to the level of linear 
perturbations. By the same token, our analysis  indicates how to stabilize those models 
even for constant $w$, by selecting for them the range and size of model parameters 
which avoids a positive, and thus catastrophic, doom factor.

\section*{Acknowledgments}
We are indebted to M. Beltr\'an and R. Ruiz de Austri for their kind help.  We also acknowledge 
R. Jim\'enez and L. Verde for very useful comments and discussions. 
The work of M.B.G., D.H. and L.L-H. was partially supported by CICYT through the project FPA2006-05423 and  
by CAM through the project HEPHACOS, P-ESP-00346. Furthermore, the work of all authors is partially supported by the PAU (Physics of the accelerating universe) Consolider Ingenio 2010. D.H. also acknowledges financial support from the spanish government 
through a FPU fellowship AP20053603. The work of L.L-H receives financial support through a postdoctoral fellowship 
of the PAU Consolider Ingenio 2010 and partially through F.M.R.S. and I.I.S.N. The work of O. M.  is financially supported by the spanish Ramon y Cajal program.
The work of S.R. was partially supported an Excellence Grant of Fondazione Cariparo.

\newpage
\appendix
\section{Growth of perturbations in strongly coupled scenarios}
\label{a:grstrongfull}
 The strong coupling regime can be characterized by 
 \begin{eqnarray}
\left|\frac{Q}{\mathcal H\rho_{de}}\right|&\gg&\left|3(1+w)\right| \,,
  \label{eq:condstr_full}\\
\left |\frac{Q}{\mathcal H\rho_{de}}\,\frac{\hat c_{s\,de}^2+1}{1+w }\right|  &\gg& \left|1-3\hat c_{s\,de}^2\right|\,,
\end{eqnarray}
which ensure that the dark-coupling terms dominate the evolution of both $\delta_{de}$ and $\theta_{de}$. 
With $c_{s\,de}^2>0$, Eq.~(\ref{eq:condstr_full}) alone is enough to define the regime.

The resulting growth equation for dark energy perturbations at large
scales is well approximated by
\begin{eqnarray}
  \label{eq:grstrongfullc}
  \delta_{de}''&\simeq&  \frac{\delta_{de}'}{a}\left(\frac Q{{\mathcal H}\rho_{de}} \,
  \frac{\hat c_{s\,de}^2+1}{1+w}+ a \left(\ln[Q/\rho_{de}]
  \right)'\right)\cr
&&+3\frac{\delta_{de}}{a^2}(\hat c_{s\,de}^2-w)
\left(\frac
    Q{{\mathcal H}\rho_{de}} \,
  \frac{1}{1+w}+ a \left(\ln[Q/\rho_{de}] \right)'\right)\cr
&&+\frac {1}{a^2{\mathcal H}}\delta[Q/\rho_{de}]
\left(\frac{\hat c_{s\,de}^2+1}{1+w}\frac Q{{\mathcal
      H}\rho_{de}}+a \left(\ln[Q/\rho_{de}] \right)'-\frac12-\frac32 w
\Omega_{de}\right)\cr
&&-\frac {1}{a {\mathcal H}}\left(\delta[Q/\rho_{de}]\right)' -(1+w)\frac{\dot h}2\,.
\end{eqnarray}
%

%
%

\bibliographystyle{hunsrt} 
\bibliography{bibdmde.bib}

\begin{thebibliography}{10}

\bibitem{Dunkley:2008ie}
J.~Dunkley et~al.
\newblock {Five-Year Wilkinson Microwave Anisotropy Probe (WMAP) Observations:
  Likelihoods and Parameters from the WMAP data}.
\newblock 2008, 0803.0586.

\bibitem{Komatsu:2008hk}
E.~Komatsu et~al.
\newblock {Five-Year Wilkinson Microwave Anisotropy Probe (WMAP)
  Observations:Cosmological Interpretation}.
\newblock 2008, 0803.0547.

\bibitem{Kowalski:2008ez}
M.~Kowalski et~al.
\newblock {Improved Cosmological Constraints from New, Old and Combined
  Supernova Datasets}.
\newblock 2008, 0804.4142.

\bibitem{Tegmark:2006az}
Max Tegmark et~al.
\newblock {Cosmological Constraints from the SDSS Luminous Red Galaxies}.
\newblock {\em Phys. Rev.}, D74:123507, 2006, astro-ph/0608632.

\bibitem{Percival:2006gt}
Will~J. Percival et~al.
\newblock {The shape of the SDSS DR5 galaxy power spectrum}.
\newblock {\em Astrophys. J.}, 657:645--663, 2007, astro-ph/0608636.

\bibitem{Caldwell:1998je}
R.~R. Caldwell, Rahul Dave, and P.~J. Steinhardt.
\newblock {Quintessential cosmology: Novel models of cosmological structure
  formation}.
\newblock {\em Astrophys. Space Sci.}, 261:303--310, 1998.

\bibitem{Zlatev:1998tr}
Ivaylo Zlatev, Li-Min Wang, and Paul~J. Steinhardt.
\newblock {Quintessence, Cosmic Coincidence, and the Cosmological Constant}.
\newblock {\em Phys. Rev. Lett.}, 82:896--899, 1999, astro-ph/9807002.

\bibitem{Wang:1999fa}
Li-Min Wang, R.~R. Caldwell, J.~P. Ostriker, and Paul~J. Steinhardt.
\newblock {Cosmic Concordance and Quintessence}.
\newblock {\em Astrophys. J.}, 530:17--35, 2000, astro-ph/9901388.

\bibitem{Wetterich:1994bg}
Christof Wetterich.
\newblock {The Cosmon model for an asymptotically vanishing time dependent
  cosmological 'constant'}.
\newblock {\em Astron. Astrophys.}, 301:321--328, 1995, hep-th/9408025.

\bibitem{Peebles:1987ek}
P.~J.~E. Peebles and Bharat Ratra.
\newblock {Cosmology with a Time Variable Cosmological Constant}.
\newblock {\em Astrophys. J.}, 325:L17, 1988.

\bibitem{Ratra:1987rm}
Bharat Ratra and P.~J.~E. Peebles.
\newblock {Cosmological Consequences of a Rolling Homogeneous Scalar Field}.
\newblock {\em Phys. Rev.}, D37:3406, 1988.

\bibitem{Carroll:1998zi}
Sean~M. Carroll.
\newblock {Quintessence and the rest of the world}.
\newblock {\em Phys. Rev. Lett.}, 81:3067--3070, 1998, astro-ph/9806099.

\bibitem{Amendola:1999qq}
Luca Amendola.
\newblock {Scaling solutions in general non-minimal coupling theories}.
\newblock {\em Phys. Rev.}, D60:043501, 1999, astro-ph/9904120.

\bibitem{Amendola:1999dr}
Luca Amendola.
\newblock {Perturbations in a coupled scalar field cosmology}.
\newblock {\em Mon. Not. Roy. Astron. Soc.}, 312:521, 2000, astro-ph/9906073.

\bibitem{Amendola:1999er}
Luca Amendola.
\newblock {Coupled quintessence}.
\newblock {\em Phys. Rev.}, D62:043511, 2000, astro-ph/9908023.

\bibitem{Comelli:2003cv}
D.~Comelli, M.~Pietroni, and A.~Riotto.
\newblock {Dark energy and dark matter}.
\newblock {\em Phys. Lett.}, B571:115--120, 2003, hep-ph/0302080.

\bibitem{Das:2005yj}
Subinoy Das, Pier~Stefano Corasaniti, and Justin Khoury.
\newblock Super-acceleration as signature of dark sector interaction.
\newblock {\em Phys. Rev.}, D73:083509, 2006, astro-ph/0510628.

\bibitem{Valiviita:2008iv}
Jussi Valiviita, Elisabetta Majerotto, and Roy Maartens.
\newblock {Instability in interacting dark energy and dark matter fluids}.
\newblock 2008, 0804.0232.

\bibitem{newUS}
M.~B. Gavela et~al.
\newblock {Work in progress}.

\bibitem{Kodama:1985bj}
Hideo Kodama and Misao Sasaki.
\newblock {Cosmological Perturbation Theory}.
\newblock {\em Prog. Theor. Phys. Suppl.}, 78:1--166, 1984.

\bibitem{Bean:2007ny}
Rachel Bean, Eanna~E. Flanagan, and Mark Trodden.
\newblock Adiabatic instability in coupled dark energy-dark matter models.
\newblock 2007, arXiv:0709.1128 [astro-ph].

\bibitem{Lifshitz:1945du}
E.~Lifshitz.
\newblock {On the Gravitational stability of the expanding universe}.
\newblock {\em J. Phys. (USSR)}, 10:116, 1946.

\bibitem{Lifshitz:1963ps}
E.~M. Lifshitz and I.~M. Khalatnikov.
\newblock {Investigations in relativistic cosmology}.
\newblock {\em Adv. Phys.}, 12:185--249, 1963.

\bibitem{Ma:1995ey}
Chung-Pei Ma and Edmund Bertschinger.
\newblock {Cosmological perturbation theory in the synchronous and conformal
  Newtonian gauges}.
\newblock {\em Astrophys. J.}, 455:7--25, 1995, astro-ph/9506072.

\bibitem{Lewis:1999bs}
Antony Lewis, Anthony Challinor, and Anthony Lasenby.
\newblock Efficient computation of {CMB} anisotropies in closed {FRW} models.
\newblock {\em Astrophys. J.}, 538:473--476, 2000, astro-ph/9911177.

\bibitem{Bean:2003fb}
Rachel Bean and Olivier Dore.
\newblock {Probing dark energy perturbations: the dark energy equation of state
  and speed of sound as measured by WMAP}.
\newblock {\em Phys. Rev.}, D69:083503, 2004, astro-ph/0307100.

\bibitem{Weller:2003hw}
Jochen Weller and A.~M. Lewis.
\newblock {Large Scale Cosmic Microwave Background Anisotropies and Dark
  Energy}.
\newblock {\em Mon. Not. Roy. Astron. Soc.}, 346:987--993, 2003,
  astro-ph/0307104.

\bibitem{Amendola:2000uh}
Luca Amendola and Domenico Tocchini-Valentini.
\newblock {Stationary dark energy: the present universe as a global attractor}.
\newblock {\em Phys. Rev.}, D64:043509, 2001, astro-ph/0011243.

\bibitem{Olivares:2006jr}
German Olivares, F.~Atrio-Barandela, and D.~Pavon.
\newblock {Matter density perturbations in interacting quintessence models}.
\newblock {\em Phys. Rev.}, D74:043521, 2006, astro-ph/0607604.

\bibitem{CalderaCabral:2008bx}
Gabriela Caldera-Cabral, Roy Maartens, and L.~Arturo Urena-Lopez.
\newblock {Dynamics of interacting dark energy}.
\newblock 2008, 0812.1827.

\bibitem{delCampo:2008jx}
Sergio del Campo, Ramon Herrera, and Diego Pavon.
\newblock {Interacting models may be key to solve the cosmic coincidence
  problem}.
\newblock 2008, 0812.2210.

\bibitem{Clarkson:2007bc}
Chris Clarkson, Marina Cortes, and Bruce~A. Bassett.
\newblock Dynamical dark energy or simply cosmic curvature?
\newblock {\em JCAP}, 0708:011, 2007, astro-ph/0702670.

\bibitem{Amendola:2007nt}
Luca Amendola and Shinji Tsujikawa.
\newblock {Phantom crossing, equation-of-state singularities, and local gravity
  constraints in $f(R)$ models}.
\newblock {\em Phys. Lett.}, B660:125--132, 2008, 0705.0396.

\bibitem{Tsujikawa:2008uc}
Shinji Tsujikawa, Kotub Uddin, Shuntaro Mizuno, Reza Tavakol, and Jun'ichi
  Yokoyama.
\newblock {Constraints on scalar-tensor models of dark energy from
  observational and local gravity tests}.
\newblock {\em Phys. Rev.}, D77:103009, 2008, 0803.1106.

\bibitem{Lewis:2002ah}
Antony Lewis and Sarah Bridle.
\newblock {Cosmological parameters from CMB and other data: a Monte- Carlo
  approach}.
\newblock {\em Phys. Rev.}, D66:103511, 2002, astro-ph/0205436.

\bibitem{Freedman:2000cf}
W.~L. Freedman et~al.
\newblock {Final Results from the Hubble Space Telescope Key Project to Measure
  the Hubble Constant}.
\newblock {\em Astrophys. J.}, 553:47--72, 2001, astro-ph/0012376.

\bibitem{Simon:2004tf}
Joan Simon, Licia Verde, and Raul Jimenez.
\newblock {Constraints on the redshift dependence of the dark energy
  potential}.
\newblock {\em Phys. Rev.}, D71:123001, 2005, astro-ph/0412269.

\bibitem{Wang:2007mza}
Yun Wang and Pia Mukherjee.
\newblock {Observational Constraints on Dark Energy and Cosmic Curvature}.
\newblock {\em Phys. Rev.}, D76:103533, 2007, astro-ph/0703780.

\bibitem{Lesgourgues:2006nd}
Julien Lesgourgues and Sergio Pastor.
\newblock {Massive neutrinos and cosmology}.
\newblock {\em Phys. Rept.}, 429:307--379, 2006, astro-ph/0603494.

\bibitem{Hannestad:2005gj}
Steen Hannestad.
\newblock {Neutrino masses and the dark energy equation of state: Relaxing the
  cosmological neutrino mass bound}.
\newblock {\em Phys. Rev. Lett.}, 95:221301, 2005, astro-ph/0505551.

\bibitem{LaVacca:2008mh}
G.~La~Vacca, S.~A. Bonometto, and L.~P.~L. Colombo.
\newblock {Higher neutrino mass allowed if DM and DE are coupled}.
\newblock 2008, 0810.0127.

\bibitem{Chevallier:2000qy}
Michel Chevallier and David Polarski.
\newblock {Accelerating universes with scaling dark matter}.
\newblock {\em Int. J. Mod. Phys.}, D10:213--224, 2001, gr-qc/0009008.

\bibitem{Linder:2002et}
Eric~V. Linder.
\newblock {Exploring the expansion history of the universe}.
\newblock {\em Phys. Rev. Lett.}, 90:091301, 2003, astro-ph/0208512.

\bibitem{newraul}
{Work in progress in collaboration with R.~Jim\'enez and L.~Verde}.

\end{thebibliography}
%
%
\end{document}